\pdfoutput=1
\documentclass[twocolumn,showpacs,amsmath,amssymb,footinbib,prb]{revtex4}
\usepackage{dsfont}
\usepackage[pdftex]{graphicx}
\usepackage[pdftex]{hyperref}
\usepackage{graphicx}
\usepackage{dcolumn}
\usepackage{bm}
\usepackage[]{color}
\usepackage[T1]{fontenc}
\usepackage{rotating}
\newcommand{\be}{\begin{equation}}
\newcommand{\ee}{\end{equation}}
\newcommand{\bea}{\begin{eqnarray}}
\newcommand{\eea}{\end{eqnarray}}
\newcommand{\bit}{\begin{itemize}}
\newcommand{\eit}{\end{itemize}}
\newcommand{\non}{\nonumber \\}
\newcommand{\nonu}{\nonumber}
\newcommand{\noi}{\noindent}

\newcommand{\upa}{\uparrow}
\newcommand{\doa}{\downarrow}
\newcommand{\ra}{\rangle}
\newcommand{\la}{\langle}
\newcommand{\rar}{\rightarrow}

\newcommand{\Sigh}{\hat{\Sigma}}
\newcommand{\ah}{\hat{a}}
\newcommand{\bh}{\hat{b}}

\begin{document}

\title{Spin entanglement using coherent light and cavity-QED}
\author{Julian Grond}
\email{julian.grond@edu.uni-graz.at}
\author{Walter P\"otz}
\affiliation{Institut f\"ur Physik, Karl-Franzens-Universit\"at Graz, Universit\"atsplatz 5, 8010 Graz, Austria}
\author{Atac Imamoglu}
\affiliation{Institut f\"ur Quantenelektronik, ETH Z\"urich, Wolfgang-Pauli-Strasse 16, 8093 Z\"urich, Switzerland}

\begin{abstract}
A scheme for probabilistic entanglement generation between two distant single electron doped quantum dots, each placed in a high-Q microcavity, by detecting strong coherent light which has interacted dispersively with both subsystems and experienced Faraday rotation due to the spin selective trion transitions is discussed. In order to assess the applicability of the scheme for distant entanglement generation between atomic qubits proposed by T. D. Ladd \textit{et al.} [New J. Phys. \textbf{8}, 184 (2006)] to two distant quantum dots, one needs to understand the limitations imposed by hyperfine interactions of the quantum dot spin with the nuclear spins of the material and by nonidentical quantum dots. Feasibility is displayed by calculating the fidelity for Bell state generation analytically within an approximate framework. The fidelity is evaluated for a wide range of parameters and different pulse lengths, yielding a trade-off between signal and decoherence, as well as a set of optimal parameters. Strategies to overcome the effect of non-identical quantum dots on the fidelity are examined and the time scales imposed by the nuclear spins are discussed, showing that efficient entanglement generation is possible with distant quantum dots. In this context, effects due to light hole transitions become important and have to be included. The scheme is discussed for one- as well as for two-sided cavities, where one must be careful with reflected light which carries spin information. The validity of the approximate method is checked by a more elaborate semiclassical simulation which includes trion formation.
\end{abstract}


\maketitle

\section{Introduction\label{sec:Int}}
An electronic spin confined in a semiconductor quantum dot (QD) is an important candidate for a potential building block of future quantum computers \cite{Los,Ima2,Tay1} due to the long relaxation and coherence times, which are measured to exceed $20\;\mathrm{ms}$ and $10\;\mu s$, respectively.\cite{Dre} Entanglement between distant electronic qubits using strong coherent light and dispersive interaction has been proposed to be useful for large distance quantum repeaters. \cite{Loo} In this scheme, direct interactions between the qubits do not play a role but the entanglement is achieved by letting both quantum systems interact with a laser pulse which acquires a phase shift conditional on the state of the qubit and in turn is measured by homodyne detection and the entanglement is distributed over kilometers. \cite{Nem,Mun} Thereby, the spin degrees of freedom are projected into a maximally entangled state. In Ref. \onlinecite{Lad}, a situation was discussed  where each electronic qubit is placed in a high-Q microcavity for better results.

In this work, we analyze feasibility of this scheme as a laboratory experiment of high-fidelity entanglement creation using the spin of an excess electron in a self-assembled QD in a cavity as qubit and exploiting the spin-selective trion transitions which lead to Faraday rotation of the light which has interacted dispersively. Faraday rotation with QDs has been measured by Atat\"ure \textit{et al}. \cite{Ata1} The main source of decoherence is due to light scattering when it interacts with the QDs, and thus we aim to analyse the fidelity \cite{Nie,Mun} for Bell state generation, including the effects of the measurement uncertainty and of decoherence. As for typical QD parameters the saturation of the interaction is rather low, it is very helpful to use a simple analytical model by eliminating the excited states, which enormously simplifies the analysis of the fidelity in terms of all the parameters involved and the identification of their optimal values. 

In order to apply the scheme of Ref. \onlinecite{Lad} to distant QDs, one has to take into account that QDs are less ideal objects than atoms and, in general, the two QDs do not have equal properties. As entanglement generation relies upon indistinguishability of the two cases where the QDs have opposite spin, it is important to understand the dependence on deviating parameters and to work out strategies to overcome this limitation. Eventually,  we also have to consider the valence band structure and,  in addition to heavy hole transitions, also take into account light hole transitions which are further detuned and couple more weakly. However, the effects are non-negligible 
for some scenarios.

As QDs interact with the nuclear spins in the solid which let the created entangled states dephase, limiting time scales are imposed which will be discussed. Saturation effects not included in the simple model are checked via semiclassical simulations. Primarily, we discuss not only one-sided cavities which can be used with acousto-optic modulators (AOM) that bring in the laser and then allow for the reflected light to go off in a different direction but also two-sided cavities which seem simpler in the sense that the light can linearly pass them. However, the reflected light lost into the environment destroys the entanglement and one has to be careful here.

First, we shortly describe trions and distant entanglement generation, motivate the use of a cavity, and describe the time scales imposed by the nuclear spins in Sec. \ref{sec:Sch}. The systems Hamiltonian, the expansion, and approximations made in order to get a simpler model, as well as the resulting dynamics, are discussed for several pulse lengths in Sec. \ref{sec:Fid}, followed by the discussion and evaluation of the fidelity with equal parameters at each QD-cavity system. The problems when using two-sided cavities and implications are presented thereafter. In Sec. \ref{sec:Non}, we go into the issue of nonidentical QDs followed by a discussion of light hole transitions and its effect on the scheme in Sec. \ref{sec:Light}. Finally, we test our model by semiclassical simulations before we draw conclusions in Sec. \ref{sec:Sum}. The appendices are concerned with details about the expression for the fidelity and details related to nonidentical QDs.

\section{Scheme and its limitations\label{sec:Sch}} 

When a gate voltage is applied such that exactly one electron tunnels in, a QD has two ground states with spin $\pm\frac{1}{2}$ ($|\upa\ra$, $|\doa\ra$) and thus also two possible excitations called trions, consisting of a of hole and two electrons with antiparallel spin in the lowest lying conduction band state ($|\upa\doa,\Uparrow\ra$, $|\upa\doa,\Downarrow\ra$) with spin $\pm\frac{3}{2}$, when only heavy hole transitions are considered. According to the optical selection rules, $|\upa\doa,\Uparrow\ra$ is excited only with left circular (or plus) polarized light, and $|\upa\doa,\Downarrow\ra$ only with right (or minus) circular polarized light (strong trion transitions). The net effect of spinflips induced by heavy-light hole mixing, leading to nonspin preserving (weak trion) transitions and coherently coupling the ground states, is as small as that due to nuclear spins at an externally applied magnetic field $B_{ext}=1\;\mathrm{T}$.\cite{Dre} 

Thus, effectively we can treat the QD as a four-level system, where light of a definite circular polarization only sees a two-level system.  The interaction with a highly detuned field is mainly dispersive and can be calculated by putting the susceptibility $\chi$ of a nonabsorptive medium \cite{Lou} in the limit of large detuning into the slowly varying envelope approximation (SVEA) equation. The phase shift acquired after interaction with the field of a laser pulse of length $L$, with cross section $A_L$ and frequency $\omega_L$, can be expressed as \footnote{We use units such that $\hbar=1$.}
\be
\theta=\frac{\eta\omega_L}{2}\epsilon_0\chi\cdot L=\frac{1}{4}\frac{\sigma_0}{A_L}\frac{\Gamma}{\Delta\omega}\;,\quad\textrm{with}\quad\chi\approx\frac{|\mu_{eg}|^2}{\epsilon_0 \Delta\omega V}\;,
\ee
where $\Delta\omega=\nu-\omega_L$ is the detuning, $\Gamma=\frac{\nu^3|\mu_{eg}|^2}{3\pi\epsilon_0 c^3}$ is the radiative decay rate, $\mu_{eg}$ is the dipole matrix element of the transition with frequency $\nu$, and $\sigma_0=\frac{3\lambda^2}{2\pi}$ is the scattering cross section of the QD with $\lambda=\frac{2\pi c}{\nu}$ and $\eta=\sqrt{\frac{\mu_0}{\epsilon_0}}$. 

When dissipation is neglected, we have an effective Hamiltonian for the laser light and QD spins,
\be
H_x=-\sum_{q=0,1}J_{xq}\ah_{q}^{\dagger}\ah_{q}|g_q\ra_x\la g_q|_x
\ee
at each subsystem $x=A,B$, where $q=1$ stands for spin up or plus polarization and $q=0$ for spin down or minus polarization, depending on the context. $g_q$ denotes the ground state of spin $q$. When using $x$-polarized light as input ($|\alpha_{\textrm{\tiny IN}}\ra_x|0\ra_y=|\frac{\alpha_{\textrm{\tiny IN}}}{\sqrt{2}}\ra_+|\frac{\alpha_{\textrm{\tiny IN}}}{\sqrt{2}}\ra_-$), we get for $J_{xq}\equiv J$, interaction time $t=\frac{\theta}{J}$ and $\theta\ll1$ a rotation of the polarization plane (Faraday rotation represented by $\hat{U}_x=e^{-iH_xt}$),
\bea\label{eq:ideal}
\hat{U}_A\hat{U}_B\frac{1}{\sqrt{2}}\bigl[|g_0\ra_A+|g_1\ra_A\bigr]  \frac{1}{\sqrt{2}}\bigl[|g_0\ra_B-|g_1\ra_B\bigr]|0\ra_y\\ 
\rightarrow \frac{1}{2}\bigl[\sum_{q=0,1} |g_q\ra_A|g_q\ra_B|(-1)^q\theta\alpha_{\textrm{\tiny IN}} \ra_y+\sqrt{2}|\Psi^-\ra|0\ra_y\bigr]\;, \nonu	
\eea
where we discard the $x$-polarized component because it does not matter. $|\Psi^-\ra=\frac{1}{\sqrt{2}}\bigl[|g_0\ra_A|g_1\ra_B-|g_1\ra_A|g_0\ra_B\bigr]$ is the Bell singlet state. Experimentally, the initial state of the QDs in Eq. \eqref{eq:ideal} can be achieved by spin-flip Raman transitions \cite{Ima} applied to a spin state prepared with high fidelity by spin pumping. \cite{Ata2}

We will show now by estimated conditions, that overcoming photon scattering requires the need of a cavity, similar to Ref. \onlinecite{Sug} where the case of a quantum nondemolition measurement using Faraday rotation was discussed. We note here that their conditions were certainly too strict as scattering is not harmful in a readout experiment. Decoherence caused by the decay of the trions with a linewidth of maximally $\Gamma=0.002\;\mathrm{meV}$ leads to elastic (\textit{Rayleigh}) scattering at rate $\Gamma\frac{g^2}{\Delta\omega^2}$ ($g=\sqrt{\frac{\omega_L}{2\epsilon_o A_L L_c}}|\mu_{eg}|$ is the light-matter coupling) at the emitter, \cite{Lou} thereby revealing the spin state. Therefore, the number of scattering processes should be kept small, basically less than $1$, while the signal-to-noise ratio (SNR) from Eq. \eqref{eq:ideal} should be greater than $1$. When a one-sided cavity is used for each QD, having the effect of enhancing the coupling by the finesse factor $\mathcal{F}=\frac{2\pi }{\kappa T_{rt}}$, where $T_{rt}=\frac{2 L_c}{c}$ is the round-trip time, $L_c$ is the length, $A_c$ is the area, and $\kappa$ is the decay rate of the cavity, we aim to fulfill the following conditions:
\bea\label{eq:estimates}
\textrm{SNR}&=&\frac{\mathcal{F}}{\pi} \frac{\sigma_0}{A_c}\frac{\Gamma\alpha_{\textrm{\tiny IN}}}{\Delta\omega}>1\;,\non
n_{scatt}&=&2\frac{\mathcal{F}}{\pi}\frac{\sigma_0}{A_c}\frac{\Gamma^2\alpha_{\textrm{\tiny IN}}^2}{\Delta\omega^2}<1\;.
\eea
For a cavity with $\mathcal{F}\sim 10^4$, there is a regime due to the, respectively, linear and quadratic dependences of SNR
and $n_{scatt}$ on $\frac{\Gamma\alpha_{\textrm{\tiny IN}}}{\Delta\omega}$, in sharp contrast to the case without a cavity. Elimination of $\alpha_{\textrm{\tiny IN}}$ yields the necessary condition,
\be\label{eq:regime}
g^2>\frac{\Gamma\kappa}{2}\;,
\ee
which corresponds to an intermediate coupling regime.

Compared to atoms, QDs are  certainly less ideal objects: the transition frequency and the strength of the light-matter interaction of two self-assembled QDs will in general never be identical as they cannot be controlled in the growing process and the QDs are chosen out of many randomly distributed samples. As our parameter space is quite large, analyzing strategies to get around this problem are helped a lot by using a simpler model than solving differential equations for each set of parameters. 

For the preservation of entanglement between distant QDs, the limiting time scales are due to hyperfine coupling with the mesoscopic bath of the nuclear spins of the lattice. For an externally applied magnetic field of the size of $B_{ext}\sim1\; T$, spin flips are largely suppressed. \cite{Dre} The effective magnetic field of the nuclei orthogonal to $B_{ext}$ can be eliminated by means of a rotating wave approximation. \cite{Tay3} In other words, spin flips are prevented by energy conservation. As the nuclear spin correlation time ($\sim1\;\mathrm{ms}$) is large compared to the timescale for entanglement generation, we may treat them in the quasistatic approximation assuming a constant nuclear ("Overhauser") field of $B_{nuc}=15\;mT$ for InAs/GaAs-QDs, \cite{Dre} which is different for each experimental run and may be treated as a classical Gaussian distributed random variable. \cite{Tay3} Thus, as singlet and triplet zero (singlet with a relative plus sign) get mixed in an unknown way due to the different $B_{nuc}^A$ and $B_{nuc}^B$ at each dot, the entangled state completely dephases at a timescale of $T_2^*=(\gamma_e B_{nuc})^{-1}\sim 1.26\;\mathrm{ns}$. However, by applying spin echo, which should be uncomplicated when using electric-dipole-induced spin resonance, \cite{Gol2} the singlet rephases after twice the time interval between the preparation of the initial state from Eq. \eqref{eq:ideal} and the spin flips: At any given time the two QD spins are in an unknown superposition of singlet and triplet.  When the entangled state is going to be used for some task at a specified time, spin echo is used to ensure that the state has rephased into a singlet. Clearly, one must keep track also of the phases when the two QDs differ in parameters and due to Zeeman splitting by $B_{ext}$. Spin echo signals, as has been measured, e.g., by Petta \textit{et. al.}, \cite{Pet} decay due to the variation of the nuclear spins at longer time scales i.e. spin coherence is lost irreversibly into the environment at $T_2\sim10\;\mu s$. \cite{Dre} 

The time scale on which entanglement can be generated is now determined by the time during which the initial state can be prepared ($t_{prep}<1\; ns$), by the propagation time ($t_{prop}\sim ns$), and by the pulse length which we will determine below. 

\section{Fidelity within an approximate model\label{sec:Fid}}

Our strategy is to first expand the Hamiltonian for a cavity containing a QD spin (four-level system) and eliminate the upper levels. From this strongly simplified Hamiltonian, we derive a Markovian master equation and, since the expansion implies discarding all anharmonic terms, treat the light classically.\\

\subsection{Hamiltonian}

The Hamiltonian for a single one-sided cavity containing a QD with one excess electron and driven by a laser pulse is obtained by making the typical approximations which are common in quantum optics for a system with several inputs and outputs and a microscopic description of system and bath, \cite{Gar}
\be\label{eq:OrigH}
H=H_0+H_{cav-bath}+H_{QD-bath}+H_{JC}\;,
\ee
\noi with
\begin{widetext}
\bea
H_0 &=& \sum_{q=0,1}\frac{\nu_q}{2}\hat{\Sigma}_{q}^z+\sum_{q=0,1}\omega_0\hat{a}_{q}^{\dagger}\hat{a}_{q}+ \sum_{q=0,1}\int d\omega\;\omega\hat{b}_{q}^{\dagger}(\omega)\hat{b}_{q}(\omega)\non
&&  + \sum_{q=0,1}\int d\omega\;\omega\hat{c}_{q}^{\dagger}(\omega)\hat{c}_{q}(\omega)\;,\non
H_{JC}&=&\sum_{q=0,1}g\bigl(\hat{\Sigma}_{q}^{+}\hat{a}_{q}+\hat{\Sigma}_{q}^-\hat{a}_{q}^+\bigr)\;,\non H_{cav-bath}&=&-i\sum_{q=0,1}\sqrt{\frac{\kappa}{2\pi}}\int d\omega \bigl(\hat{a}_{q}^{\dagger}\hat{b}_{q}(\omega)-\hat{a}_{q}\hat{b}_{q}^{\dagger}(\omega)\bigr)
\;,\label{eq:cav-bath}\non
H_{QD-bath}&=&-i\sum_{q=0,1}\sqrt{\frac{\Gamma}{2\pi}}\int d\omega\bigl(\hat{\Sigma}_{q}^{+}\hat{c}_{q}(\omega)-\hat{\Sigma}_{q}^-\hat{c}_{q}^{\dagger}(\omega)\bigr)\;.\label{eq:QD-bath}
\eea
\end{widetext}
 The cavity mode operators $\hat{a}_q$  with energy $\omega_0$ are coupled with coupling constant $\sqrt{\kappa}$ to Markovian reservoirs described by continuum operators $\hat{b}_{q}(\omega)$. We denote the four states of the QD  as $|g_{q=0,1}\ra$ for the ground states and $|e_{q=0,1}\ra$ for the excited states (trions), separated by an energy $\nu$. Including the Zeeman shifts we have transition frequencies $\nu_q=\nu-(-1)^q (g_h-g_e)\mu_B B_{ext}+(-1)^q g_e \mu_B B_{nuc}$, where the electron and hole g factors are $-0.6$ and $1.8$, respectively, and the nuclear spins only couple to the electrons. As $B_{nuc}\ll B_{ext}$, its effect on the phase shift is negligible. The operators describing this four-level system are $\Sigma_{q}^+=|e_q\ra\la g_q|,\;\Sigma_{q}^-=|g_q\ra\la e_q|$ and $\Sigma_{q}^z=|e_q\ra\la e_q|-|g_q\ra\la g_q|$. These are coupled to reservoir operators $\hat{c}_q(\omega)$ with coupling constant $\sqrt{\Gamma}$ and to the cavity fields within the Jaynes-Cummings model \cite{Cum} with $g$. 

\subsection{Expansion\label{subsec:Exp}}

A systematic expansion for large detuning is achieved by applying a Schrieffer-Wolff transformation \cite{Sch} to Eq.~\eqref{eq:OrigH} 
\be
\bar{H}=e^A H e^{-A}\approx H+[A,H]+\frac{1}{2}[A,[A,H_0]]+\mathcal{O}\bigl(\frac{g^3}{\Delta\omega_q^2}\bigr)\quad,
\ee
with $A=\sum_{q=0,1}\frac{g}{\Delta\omega_{q}}\bigl(\hat{\Sigma}_{q}^+\hat{a}_{q}-\hat{\Sigma}_{q}^-\hat{a}_{q}^{\dagger}\bigr)\;$ and detuning $\Delta\omega_{q}:=\nu_q-\omega_0$ (or $\Delta\omega:=\nu-\omega_0$ without B field). Typically, the coupling constant $g$ is at least $1$ order of magnitude smaller than the detuning and the expansion is an excellent approximation, provided that $\frac{g^2 \la \ah_q^\dagger\ah_q\ra}{\Delta\omega_q^2}\ll1$. The interaction term between QDs and cavity light fields is transformed away in first order due to $[A,H_0]=-H_{JC}$ and the ideal interaction, i.e., Faraday rotation, is contained in
\bea\label{eq:Hfr}
&&H_{FR}=\frac{1}{2}[A,H_{JC}] \\
&&=\frac{1}{2}\sum_{q=0,1}\frac{g^2}{\Delta\omega_{q}}[\hat{P}_q+\hat{\Sigma}_{q}^z(1+2\ah_q^{\dagger}\ah_q)]\;, \nonu
\eea
consisting of Lamb and Stark shifts. We defined projectors onto the subspaces of given spin $q$ $\hat{P}_q=|e_q\ra \la e_q|+|g_q\ra \la g_q|$. The resulting Hamiltonian is then 
\bea\label{eq:TransfH}
&&\bar{H} = H_0+H_{FR}+H_{cav-bath}+H_{QD-bath}\\
&&+H_{Purcell}+H_{Rayleigh}+\mathcal{O}\bigl(\frac{g^3}{\Delta\omega_q^2}\bigr)\;, \nonu
\eea
The last two contributions of $\bar{H}$ lead to additional decay of the cavity fields and the trions via the interaction,
\bea
&&H_{Purcell}:=[A,H_{cav-bath}]=\\
&&-i\sum_{q=0,1}\sqrt{\frac{\kappa}{2\pi}}\frac{g}{\Delta\omega_{q}}\int d\omega(\hat{\Sigma}_{q}^{+}\hat{b}_{q}(\omega)-\hat{\Sigma}_{q}^{-}\hat{b}_{ q}^{\dagger}(\omega))\nonu
\eea
describes the \textit{Purcell} effect \cite{Ger} and leads to driving of the trion transitions because of the coherent excitation of the reservoir modeled by the $\hat{b}_{q}(\omega)$ fields. Due to the large detuning between the driving field and the trion transitions, the population in the excited state is very low and we neglect that term. This is the approximation which renders the transformed Hamiltonian particularly simple because the excited states can be completely eliminated. Rayleigh scattering is described by
\small
\bea 
H_{Rayleigh}&=&[A,H_{QD-bath}]\\
&=&-i \sum_{q=0,1}\sqrt{\frac{\Gamma}{2\pi}}\frac{g}{\Delta\omega_{q}}\hat{\Sigma}_{q}^z\int d\omega(\hat{a}_{q}^{\dagger}\hat{c}_{q}(\omega)-\hat{a}_{q}\hat{c}_{q}^{\dagger}(\omega))\;,\nonu
\eea
\normalsize
and provides the main decoherence process, as will be discussed in Sec. \ref{sec:Sch}. 

The master equation for the density operator $\hat{\rho}$ in an interaction picture with respect to $H_0$ is obtained after elimination of the excited states by making the Born-Markov approximation, common in quantum optics, \cite{Mil} and by discarding 
fast rotating terms proportional to $ e^{\pm i\Delta\omega_q t}$, such as those where $H_{cav-bath}$ and $H_{Rayleigh}$ are mixed,

\bea\label{eq:me}
\frac{d\hat{\rho}}{dt}&=&i\sum_{q=0,1}\frac{g^2}{\Delta\omega_q}\bigl[\ah_q^{\dagger}\ah_q|g_q\ra\la g_q|,\hat{\rho}\bigr]\\
&&-\sqrt{\kappa} F_{\textrm{\tiny IN}}(t) \frac{\alpha_{\textrm{\tiny IN}}}{\sqrt{2}}\sum_{q=0,1}\bigl[\ah_q-\ah_q^{\dagger},\hat{\rho}\bigr]\non
&&-\frac{\kappa}{2}\sum_{q=0,1}\bigl(\ah_q^{\dagger}\ah_q\hat{\rho}+\hat{\rho}\ah_q^{\dagger}\ah_q-2\ah_q\hat{\rho}\ah_q^{\dagger}\bigr)\non
&&+\sum_{q=0,1}\Bigl(\hat{L}_q\hat{\rho}\hat{L}_q^{\dagger}-\frac{1}{2}\{\hat{L}_q^{\dagger}\hat{L}_q,\hat{\rho}\}\Bigr)\;,\nonu
\eea
where the first term accounts for the Stark shift, the second and third ones describe a driven damped cavity and the last one Rayleigh scattering at rate $\Gamma_q^R=\frac{g^2 \Gamma}{\Delta\omega_q^2}$ with Lindblad operator $\hat{L}_{q}=\sqrt{\Gamma_q^R}\hat{a}_q|g_q\ra\la g_q|$, i.e., spin-dependent light scattering which eventually leads to the decay of the coherences. The driving field required for Faraday rotation is an $x$-polarized driving laser pulse centered at $t_0$ and is given the shape 
\be
F_{\textrm{\tiny IN}}(t)=\frac{e^{-\frac{(t-t_0)^2}{4\tau_{\textrm{\tiny P}}^2}}}{\sqrt{\sqrt{2\pi}\tau_{\textrm{\tiny P}}}}\;,
\ee
as well as photon amplitude $\alpha_{\textrm{\tiny IN}}$. This means, each circular polarization carries a number of photons $\frac{\alpha_{\textrm{\tiny IN}}}{\sqrt{2}}$. Its pulse length is $\tau_{\textrm{\tiny P}}$ and its central frequency is $\omega_L$. 
This dynamics clearly implies a classical evolution of the light (for the mean values, $\la|g_q\ra \la g_q|\ah_q^{\dagger}\ah_q\ra=\la|g_q\ra \la g_q|\ra\la\ah_q^{\dagger}\ah_q\ra$ certainly holds since $|g_q\ra \la g_q|$ is constant in time), i.e., all terms that would lead to quantum corrections to the light are in higher order in the expansion parameter. 

\subsection{Cavity fields and signal\label{subsec:Cav}}

Making an ansatz for the light of either circular polarization in terms of coherent states, the density operator for the entangled atom-cavity system is given as 
\bea\label{eq:rho}
\hat{\rho}(t)&=&\sum_{q=0,1}\Bigl(\rho_{g_q g_q}(t)|g_q\ra\la g_q|\otimes|\tilde{\alpha}_q(t)\ra_q\la\tilde{\alpha}_q(t)|_q\\ &&\otimes|\alpha_{q'}(t)\ra_{q'}\la\alpha_{q'}(t)|_{q'}\Bigr)+\sum_{q=0,1}\Bigl(\rho_{g_{q'} g_q}(t)|g_q\ra\la g_{q'}|\non
&&\otimes|\tilde{\alpha}_q(t)\ra_q\la\alpha_q(t)|_q\otimes|\alpha_{q'}(t)\ra_{q'}\la\tilde{\alpha}_{q'}(t)|_{q'}\Bigr)\;,\nonu
\eea
where $q'$ denotes the  polarization opposite to $q$.  $\tilde{\alpha}_q(t)$ and $\alpha_q(t)$, respectively, stand for the amplitudes of the coherent states when the QD is in the interacting and noninteracting spin.

The equations of motion for $\alpha_q(t)$ and $\tilde{\alpha}_q(t)$ are given as the derivative of the cavity field mean value $\la \ah\ra=Tr \bigl\{\ah \hat{\rho}\bigr\}$ for the cases $\rho_{g_{q'} g_{q'}}(0)=1$ and $\rho_{g_q g_q}(0)=1$, respectively (note that these density matrix elements are constant in time as we eliminated the excited states) using the master equation [Eq. \eqref{eq:me}]. Defining $\frac{\alpha_{\textrm{\tiny IN}}}{\sqrt{2}}S(t):=\alpha_q(t)$ for an empty cavity, the cavity field has, when driven on resonance, the shape (here and in the following, we will always neglect the damping of the light due to Rayleigh scattering as in the regimes of interest this effect is negligible)
\bea\label{eq:St}
S(t)&=&\sqrt{\kappa}e^{-i\omega_0 t}\int_0^t dt'e^{(\kappa/2)(t'-t)}F_{\textrm{\tiny IN}}(t')\non
&\approx& \frac{2}{\sqrt{\kappa}}e^{-i\omega_0 t}F_{\textrm{\tiny IN}}(t)\;,
\eea
which is the solution for a driven harmonic oscillator and the approximation holds in steady state ($\tau_{\textrm{\tiny P}}\gg\frac{1}{\kappa}$). Going on to the case where the cavity field interacts with a QD, $\frac{\alpha_{\textrm{\tiny IN}}}{\sqrt{2}}\tilde{S}_q(t):=\tilde{\alpha}_q(t)$ is given on resonance by
\bea\label{eq:Stt}
\tilde{S}_q(t)&=&\sqrt{\kappa}e^{-i\omega_0 t}\int_0^t dt'e^{(-i(g^2/\Delta\omega_q)+(\kappa/2))(t'-t)}F_{\textrm{\tiny IN}}(t')\non
&\approx& \frac{\sqrt{\kappa}}{\frac{\kappa}{2}-i\frac{g^2}{\Delta\omega_q}}F_{\textrm{\tiny IN}}(t)e^{-i\omega_0 t}\;.
\eea

\begin{figure}[h]
\begin{tabular}{l}
(a)\\
\includegraphics[width=7cm]{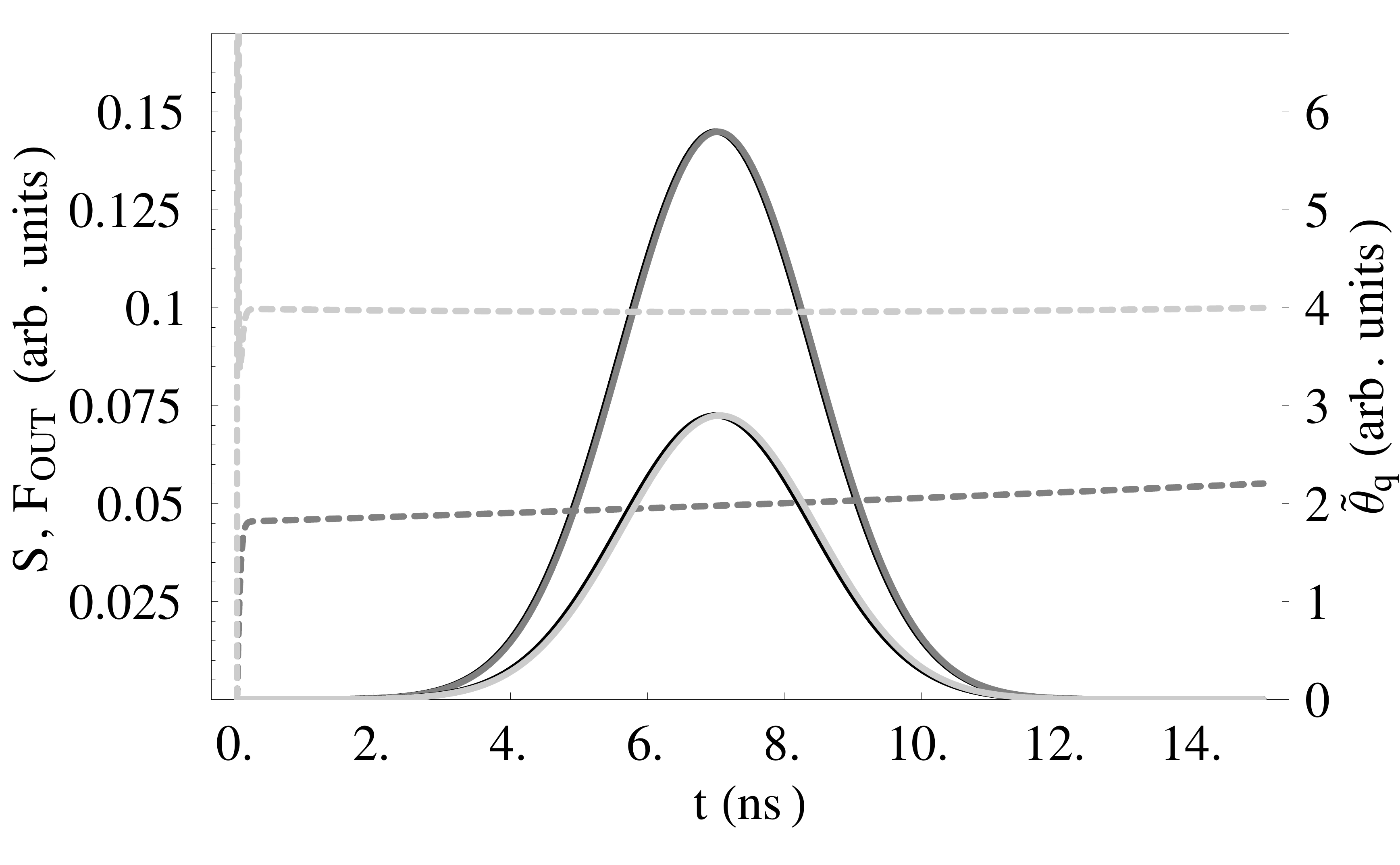}\\
(b)\\
\includegraphics[width=7cm]{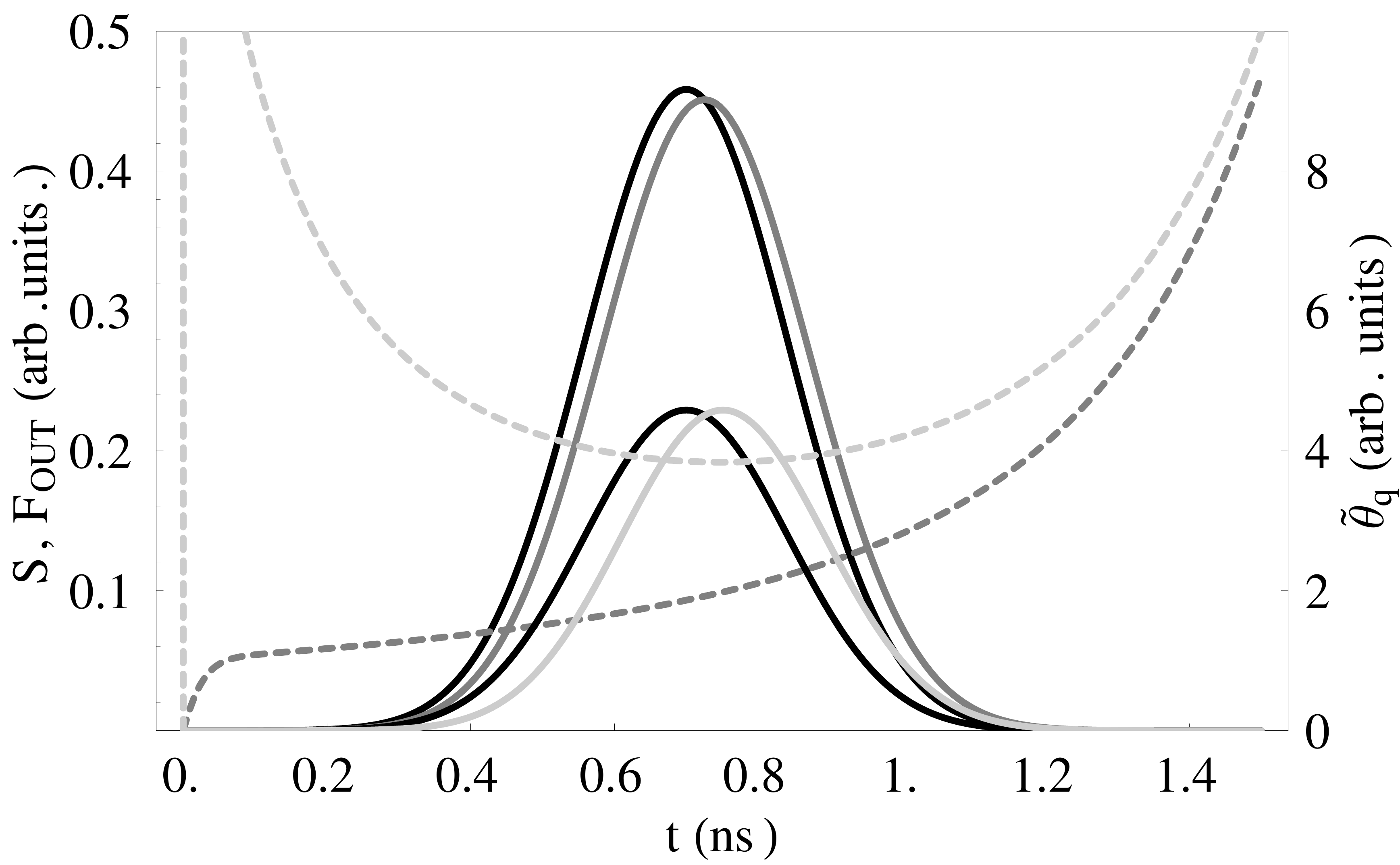}\\
(c)\\
\includegraphics[width=7cm]{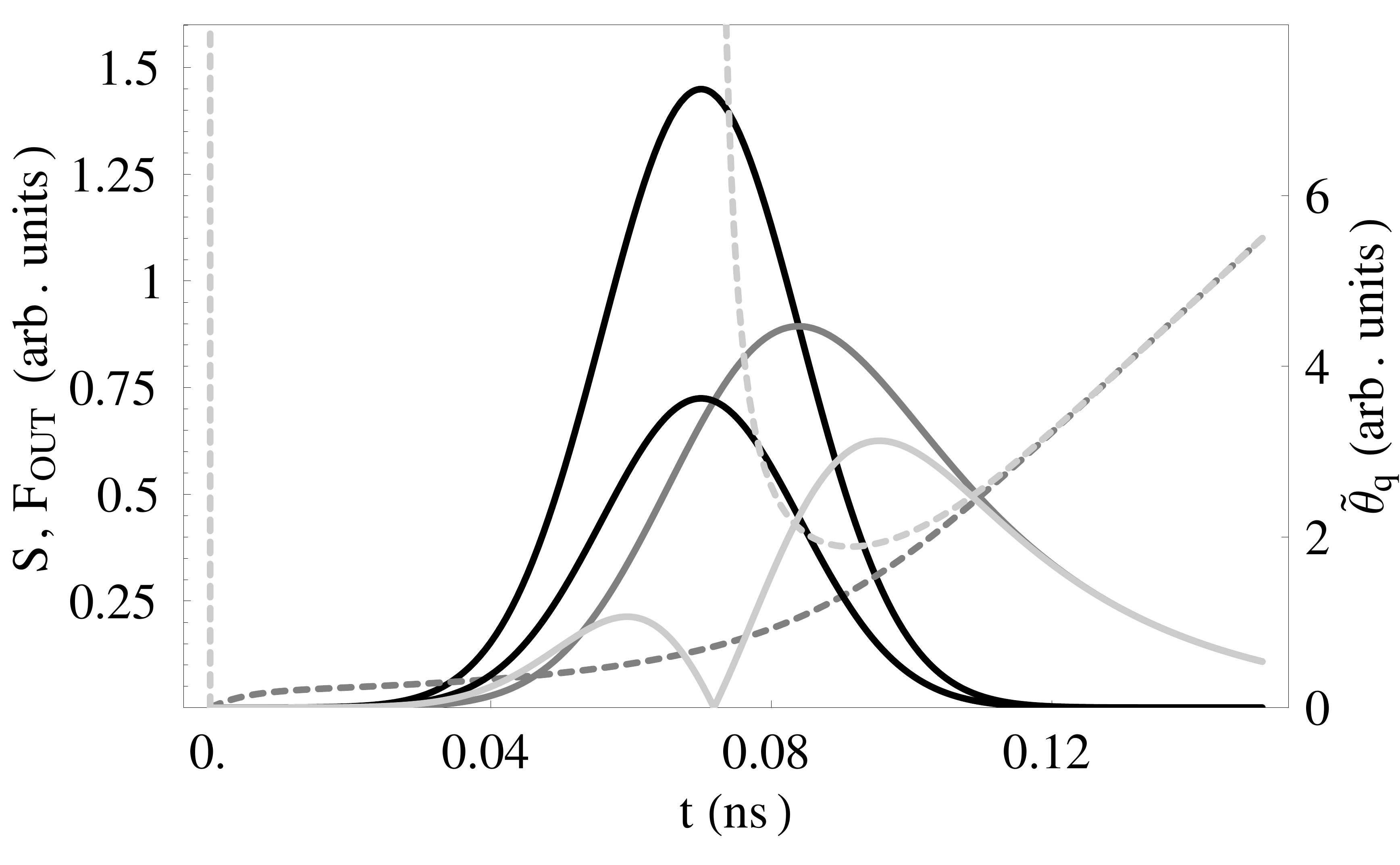}
\end{tabular}
\caption{\label{fig:cavity}$|S(t)|$ (empty cavity) for $\kappa=0.05\;\mathrm{meV}$ (gray solid line) and output field [Eq. \eqref{eq:bout}] in units, where $\kappa=1$ (light gray solid line) compared to the steady state estimates (black solid line) for pulse lengths of (a) $1\;\mathrm{ns}$, (b) $100\;\mathrm{ps}$, and (c) $10\;\mathrm{ps}$. The dashed lines illustrate the phase shift in terms of $\frac{g^2}{\kappa\Delta\omega}$. The parameters are $g=0.15\;\mathrm{meV}$, $\Delta\omega=5\;\mathrm{meV}$, and $\alpha_{\textrm{\tiny IN}}=8$. For $\tau_{\textrm{\tiny P}}=10\;\mathrm{ps}$ the rapid change of the phase corresponds to a sign change.
}
\end{figure}

The output field is related to the cavity field and the input field by means of the relation $\hat{b}_{\textrm{\tiny OUT}}^q(t)=\sqrt{\kappa} \hat{a}_q(t)-\hat{b}_{\textrm{\tiny IN}}^q(t)$ (Ref. \onlinecite{Gar}) and is a continuum field operator \cite{Blo} with the structure 
\be\label{eq:bout}
\hat{b}_{\textrm{\tiny OUT}}^q(t)=F_{\textrm{\tiny OUT}}^q(t)e^{-i\omega_L t}\ah_{\textrm{\tiny OUT}}^q\;,
\ee
whereby $F_{\textrm{\tiny OUT}}^q(t)=|\sqrt{\kappa}\tilde{S}_q(t)-F_{\textrm{\tiny IN}}(t)|$ is the pulse shape with $\int_{-\infty}^{\infty} dt |F_{\textrm{\tiny OUT}}^q(t)|^2=1$. The corresponding eigenstates are the coherent states, 
\be\label{eq:contop}
|\alpha_{\textrm{\tiny OUT}}^q(t)\ra=\exp{\Bigl[\int dt [\alpha_{\textrm{\tiny OUT}}^q(t)\hat{b}_{\textrm{\tiny OUT}}^{q,\dagger}(t)-H.c.]\Bigr]}|0\ra_q\;,
\ee
where the integral is over the real axes. For the case of dispersive interaction, we have 
\bea\label{eq:sigOS}
\alpha_{\textrm{\tiny OUT}}^q(t)=\frac{\alpha_{\textrm{\tiny IN}}}{\sqrt{2}}F_{\textrm{\tiny OUT}}^q(t)e^{i\tilde{\theta}_q(t)}\;,
\eea
which corresponds to the Faraday rotation to be measured and is equal to the estimation from Sec. \ref{sec:Sch}. In order to check the validity of the steady state assumption, we calculate Eq. \eqref{eq:St} by solving the differential equation fulfilled by $S(t)$ for several pulse lengths with results plotted in Fig. \ref{fig:cavity} and compare to the steady state curves. The phase shift is plotted in terms of $\frac{g^2}{\kappa\Delta\omega}$, corresponding to the phase shifts for large detuning.  For short pulses, the shape gets deformed and translated and we encounter nonconstant arguments for short pulse length ($\tau_{\textrm{\tiny P}} < 1\;\mathrm{ns}$). However, it is always possible to replace the nonconstant phase $\tilde{\theta}_q(t)$ in Eq. \eqref{eq:contop} by a mean phase defined as
\be\label{eq:nonc}
e^{i\bar{\theta}_q}:=\int dt |F_{\textrm{\tiny OUT}}^q(t)|^2 e^{i\tilde{\theta}_q(t)}\;.
\ee
By carrying out the integral in Eq. \eqref{eq:contop}, we can equivalently denote the output states as $|\frac{\alpha_{\textrm{\tiny IN}}}{\sqrt{2}}e^{i\bar{\theta}_q}\ra_q $, i.e., by the photons carried by the pulse. As can be seen in Fig. \ref{fig:cavity}, for pulse lengths not shorter than $\tau_{\textrm{\tiny P}}=100\;\mathrm{ps}$, the steady state approximation is quite good. Also, the output pulse shape does not depend on $q$ and we discard that index. This allows a simple decomposition into linear polarized components as
\bea
\hat{b}_{\textrm{\tiny OUT}}^q(t)&=&F_{\textrm{\tiny OUT}}(t)e^{-i\omega_L t}\frac{1}{\sqrt{2}}(\ah_{\textrm{\tiny OUT}}^x+i(-1)^q \ah_{\textrm{\tiny OUT}}^y)\non
&\equiv&\frac{1}{\sqrt{2}}(\hat{b}_{\textrm{\tiny OUT}}^x(t)+i(-1)^q\hat{b}_{\textrm{\tiny OUT}}^y(t))\;.
\eea
Balanced homodyne detection offers a means to measure the quadrature operators of the electromagnetic output field by integrating the field of interest with a large state (local oscillator) at a beam splitter and subtracting the photocurrents produced by the output. \cite{Scu} Here we consider a situation where a light pulse of $45^{\circ}$ linear polarization is split into a $y$-polarized component, which serves as the (classical) local oscillator state $\alpha_{\textrm{\tiny LO}}(t)=\alpha_{\textrm{\tiny IN}}F_{\textrm{\tiny IN}}(t)e^{-i\omega_L t}$, and an $x$-polarized component, which interacts with the QD-cavity system whereby a $y$-polarized component may be acquired. The $x$ quadrature of the latter is to be measured and thus the $x$-polarized component is removed at a polarizing beam splitter. The observable representing the homodyne detector \cite{Blo} is then given by 
\bea
\delta\hat{i}&=&q_{h}\int dt[(\bh_{\textrm{\tiny OUT}}^{y}(t))^{\dagger}\alpha_{\textrm{\tiny LO}}(t)+ H.c.]\\
&=&q_{h}2\alpha_{\textrm{\tiny IN}}\Bigl(\int dt F_{\textrm{\tiny IN}}(t)F_{\textrm{\tiny OUT}}(t)\Bigr)\hat{x}_{\textrm{\tiny OUT}}^y\;,\nonu
\eea
where $q_{h}$ is a constant related to the measurement apparatus and the integral is over the entire pulse duration. Thus at the homodyne detector, the density matrix in Eq. \eqref{eq:rho}, which can equivalently be written in terms of output fields instead of cavity fields, is projected into eigenstates of $\hat{x}_{\textrm{\tiny OUT}}^y=\frac{1}{2}(\ah_{\textrm{\tiny OUT}}^y+(\ah_{\textrm{\tiny OUT}}^y)^{\dagger})$. The factor in front of $\hat{x}_{\textrm{\tiny OUT}}^y$, in particular, $\int dt F_{\textrm{\tiny IN}}(t)F_{\textrm{\tiny OUT}}(t)$, does not matter as long as it is not too small like for pulse lengths $\tau_{\textrm{\tiny P}}<\frac{1}{\kappa}$ (see Fig. \ref{fig:cavity}). Then, clearly the relative noise is not solely determined by the variance of $\hat{x}^y$, but other contributions become important which have been neglected due to the large interference between signal pulse and local oscillator. \cite{Scu} We will thus restrict ourselves to pulse lengths $\tau_{\textrm{\tiny P}}=1\;\mathrm{ns}$ and $100\;\mathrm{ps}$ where $\tau_{\textrm{\tiny P}}\kappa\sim 70$ and $7$, respectively. Alternatively, the local oscillator pulse could be sent through the same but empty cavity structure as the signal pulse, then having a very similar shape and improved overlap.\\ 

Using pulses with $\tau_{\textrm{\tiny P}}\sim100\;\mathrm{ps}$,  one may create entanglement on a time scale of $\sim10\;\mathrm{ns}$: The interaction of the light with one cavity lasts for about $\sim 1 \;\mathrm{ns}$, the light travels typically $1\;\mathrm{ns}$ between the cavities, and after the interaction, a spin-echo pulse can be applied which rephases the desired Bell state after twice the interaction time.

\subsection{Decay of coherences}

The equation describing the coherence
\bea
\rho_{g_1 g_0}(t)&=&Tr \Bigl(|g_1\ra\la g_0|\otimes|\tilde{\alpha}_1(t)\ra_1\la\alpha_1(t)|_1\\
&&\otimes|\alpha_{0}(t)\ra_{0}\la\tilde{\alpha}_{0}(t)|_{0}\hat{\rho}(t)\Bigr)\nonu
\eea
is according to Eq. \eqref{eq:me} given by
\bea\label{eq:cohEOM}
\frac{d\rho_{g_1g_0}(t)}{dt}&=&ig^2(\frac{1}{\Delta\omega_1}-\frac{1}{\Delta\omega_0})(\alpha_{\textrm{\tiny IN}}^2/2)|S(t)|^2\rho_{g_1g_0}(t)\non
&&-(\Gamma_0^R+\Gamma_1^R/2)\frac{\alpha_{\textrm{\tiny IN}}^2}{2}|S(t)|^2\rho_{g_1g_0}(t)\;,
\eea
while the diagonal elements of $\hat{\rho}(t)$ are constant. We replaced $\tilde{\alpha}_{q}(t)$ by $\alpha_{q}(t)$ since the terms are already $\mathcal{O}\bigl(\frac{g^2}{\Delta\omega}\bigr)$. This equation describes, besides trivial phases due to Zeeman splitting (an additional phase should be added because magnetic fields also lead to a relative energy bias between the two trion transitions but we will neglect that in the following), the decay of coherent superposition states caused by Rayleigh scattering. Integration yields for the modulus
\bea\label{eq:deccoh}
|\rho_{g_1g_0}(T_{end})|&=&|\rho_{g_1g_0}(0)|e^{-\frac{\alpha_{\textrm{\tiny IN}}^2}{2}\sum_{q=0,1} \frac{\Gamma^R_q}{2}\Phi}\non
&\approx&|\rho_{g_1g_0}(0)|e^{-\sum_q n_{scatt}^q}\;,
\eea
taken at $T_{end}$ sufficiently long after the interaction such that no scattering occurs anymore. $\Phi=\int_{-\infty}^{\infty} dt |S(t)|^2$ is the pulse area, describing how many photons couple into the cavity. The decay in steady state, when according to Eq. \eqref{eq:St} $\Phi=\frac{4}{\kappa}$, is determined by the estimated number of scattered photons $n_{scatt}^q=\frac{\alpha_{\textrm{\tiny IN}}^2}{2}\frac{4 g^2 \Gamma}{ \kappa\Delta\omega_q^2}$ which is equivalent to the expression Eq. \eqref{eq:estimates} in Sec. \ref{sec:Sch}. 

For two one-sided cavities, each containing a QD, the decay is determined by the sum of all contributions from each cavity $x$ with spin $q$,
\be\label{eq:decent}
|\rho_{g_0g_1;g_1g_0}(T_{end})|=|\rho_{g_0g_1;g_1g_0}(0)|e^{-\frac{\alpha_{\textrm{\tiny IN}}^2}{2}\sum_{xq}\frac{\Gamma_{xq}^R}{2}\Phi_x}\;.
\ee
In this density matrix element, which is $- \frac{1}{4}$ for a Bell singlet and $0$ for a product state, the first and third indeces refer to the first subsystem and the others to the second one.

\subsection{Fidelity\label{subsec:Fidelity}}

When the second cavity is driven by the output from the first cavity, the signal has $y$ components given by 
\bea\label{eq:signal1}
&&|d_{11}=-\frac{\alpha_{\textrm{\tiny IN}}}{2}\sin{(\theta_{A1}+\theta_{B1})}\ra_y\;,\non
&&|d_{00}=\frac{\alpha_{\textrm{\tiny IN}}}{2}\sin{(\theta_{A0}+\theta_{B0})}\ra_y\;,\non
&&|d_{10}=-\frac{\alpha_{\textrm{\tiny IN}}}{2}(\sin{\theta_{A1}}-\sin{\theta_{B0}})\ra_y\;,\non
&&|d_{01}=\frac{\alpha_{\textrm{\tiny IN}}}{2}(\sin{\theta_{A0}}-\sin{\theta_{B1}})\ra_y\;\;.
\eea
We refer to the amplitudes $d_{11}$ (both spins up), $d_{00}$ (both spins down), $d_{10}$ (A: spin up and B: spin down), and $d_{01}$ (A: spin down and B: spin up) as \textit{distinguishabilities}. \cite{Loo} They are the centers of the corresponding probability distributions 
\be\label{eq:gauss}
G_{q_1q_2}(x):=\la x^y_{\textrm{\tiny OUT}}|_y|d_{q_1q_2}\ra_y=\bigl(\frac{2}{\pi}\bigr)^{\frac{1}{4}}e^{-(x-d_{q_1q_2})^2}\;,
\ee
How close to zero $d_{10}$ and $d_{01}$ will be determined by how well the QD cavity parameters for the two subsystems match. $d_{00}$ and $d_{11}$ are on the order of $1$ and mark the unwanted situation of parallel spins. The phase shift at QD $x$ and for polarization $q$ is given approximately by $\theta_{xq}=\frac{4 g^2}{\kappa_x\Delta\omega_{xq}}$. The more complete expressions which will be used in the following are given in Appendix \ref{app:B}.

After measurement, the density operator for the QDs and output fields depends on the outcome of the measurement ($x$) and is given as
\bea\label{eq:rhox}
\hat{\rho}(x,t)&=&\sum_{q_1,q_2,q_3,q_4=0,1}\Bigl(\rho_{g_{q_3}g_{q_4};g_{q_1}g_{q_2}}(t)\cdot|g_{q_1}g_{q_2}\ra\non
&&\times\la g_{q_3}g_{q_4}|G_{q_1q_2}(x)G_{q_3q_4}(x)\Bigl)\;.
\eea
The fidelity for a singlet is defined as 
\be\label{eq:fid2}
F=\frac{\int_{-x_c}^{x_c}dx\la\Psi^-|\hat{\rho}(x,T_{end})|\Psi^-\ra}{Tr_{\textrm{\tiny QD}}\int_{-x_c}^{x_c}dx \hat{\rho}(x,T_{end})}\;,
\ee 
normalized by the \textit{success probability}. \cite{Loo} The evaluated expressions are given in Appendix \ref{app:A}. Obviously, a \textit{measurement window} $x_c$ must be chosen to account for the overlaps of the Gaussian peaks, which introduce an uncertainty in the measurement result, by defining an interval around $x=0$ within which the measurement outcome is accepted and outside of which it is discarded. The integrals over the Gaussian functions in Eq. \eqref{eq:rhox} are given by error functions and the diagonal density matrix elements of the initial product state Eq. \eqref{eq:ideal} are constant in time at $\frac{1}{4}$ and $\rho_{g_1g_0;g_0g_1}(0)=\rho_{g_0g_1;g_1g_0}(0)=-\frac{1}{4}$.

\subsection{Results for identical quantum dots and cavities\label{subsec:Res}}

\begin{figure}[h]
\begin{tabular}{l}
(a)\\
\includegraphics[width=7cm]{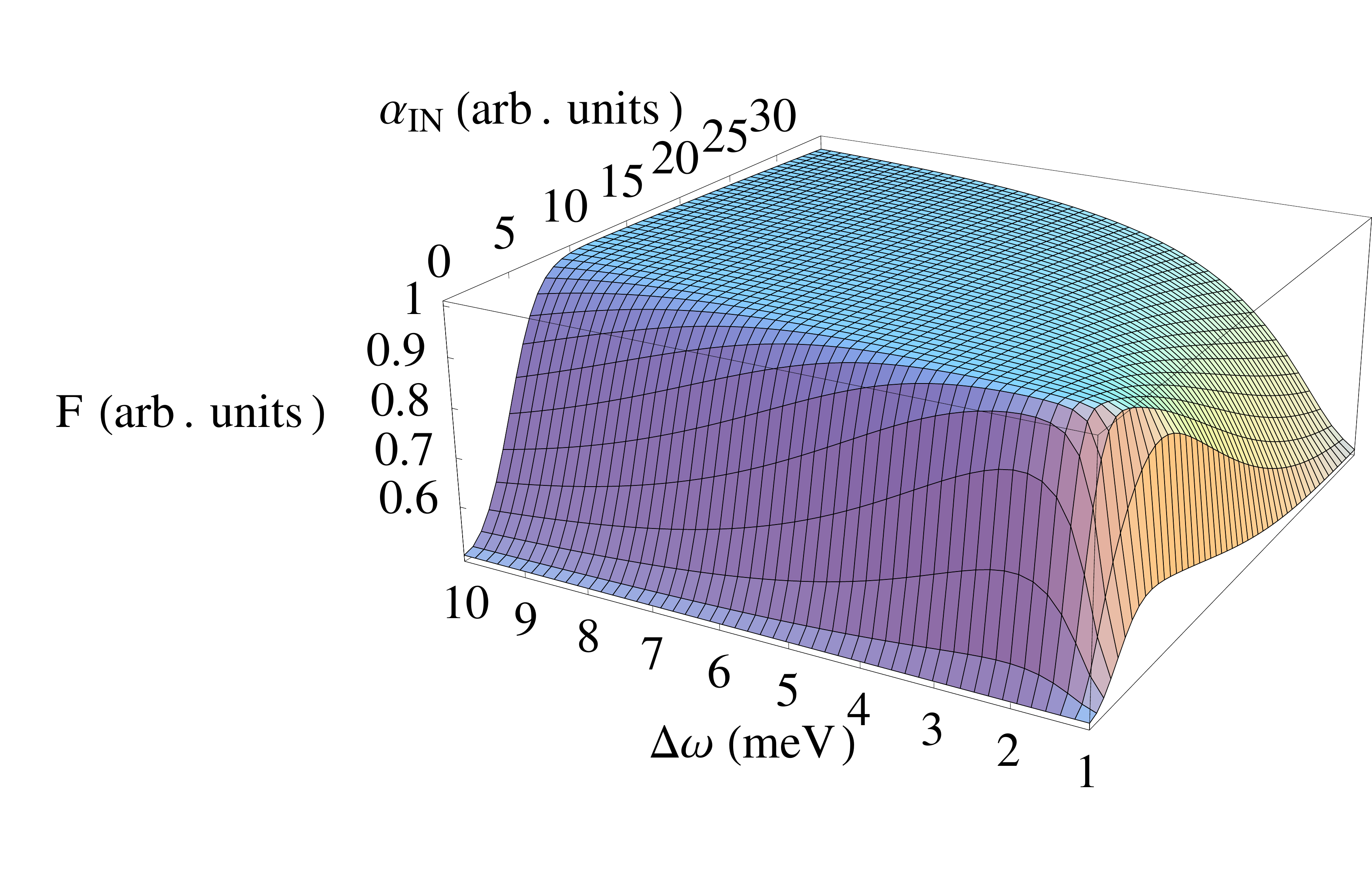}\\
(b)\\
\includegraphics[width=7cm]{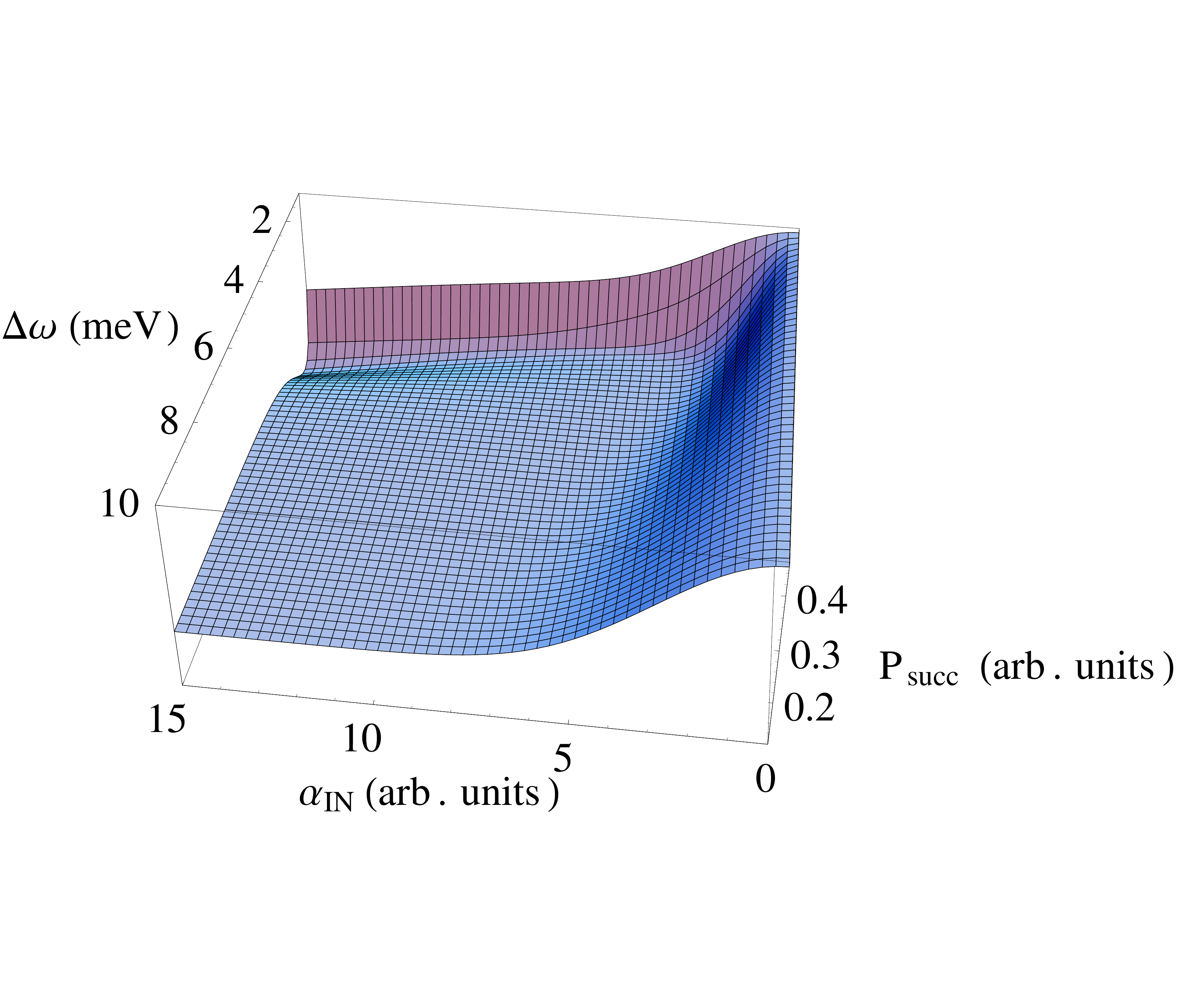}
\end{tabular}
\caption{\label{fig:fid1} (Color online) (a) Fidelity for identical parameters at each subsystem $g=0.15\;\mathrm{meV}$ and $\kappa=0.05\;\mathrm{meV}$ versus $\alpha_{\textrm{\tiny IN}}$ and $\Delta\omega$. A trade-off between decoherence and signal gives rise to an optimal set of parameters. (b) Success probability $P_s$ corresponding to the fidelity in (a), never lower than $25\%$.} 
\end{figure}

\begin{figure}[h]
\begin{tabular}{l}
(a)\\
\includegraphics[width=7cm]{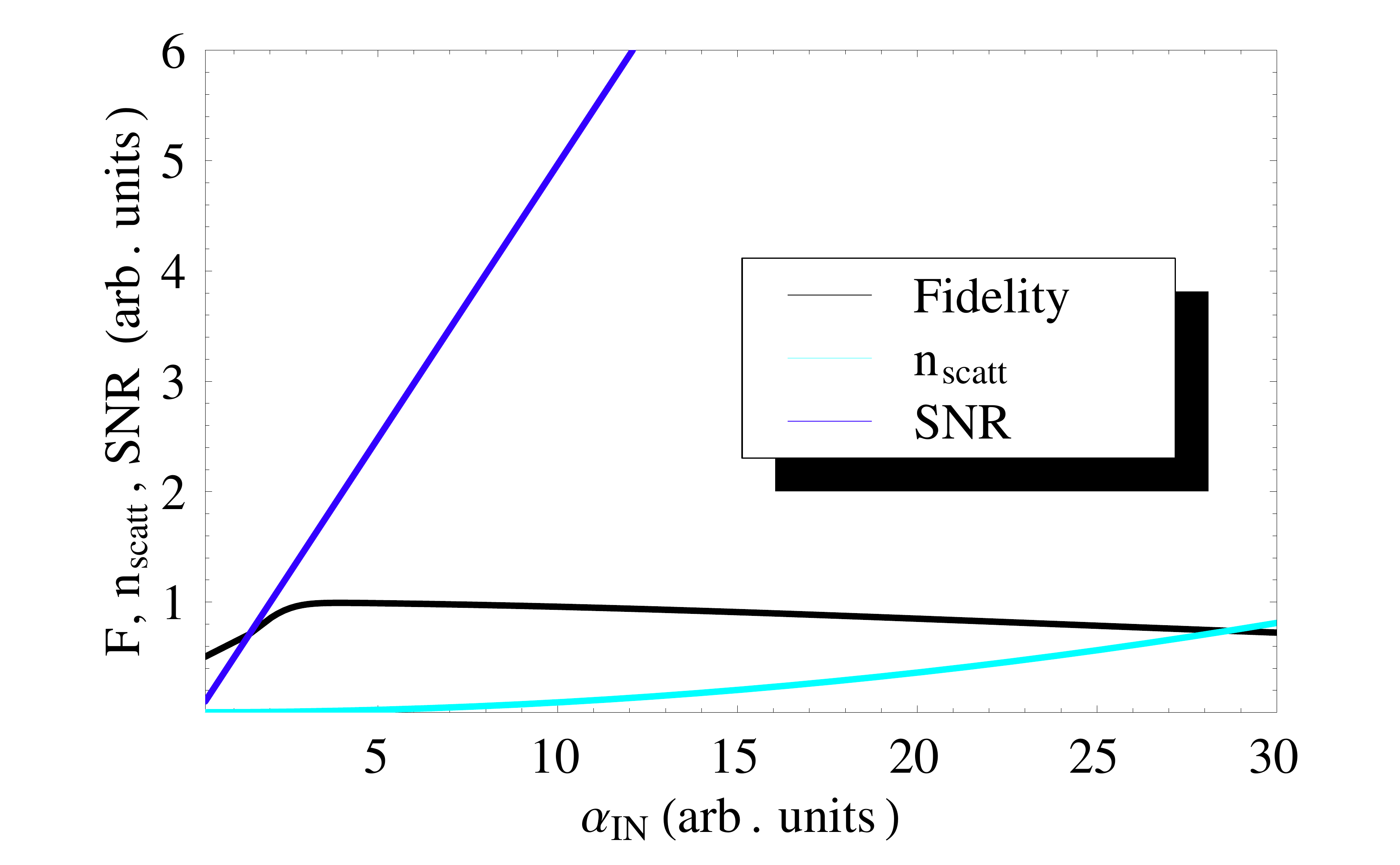}\\
(b)\\
\includegraphics[width=7cm]{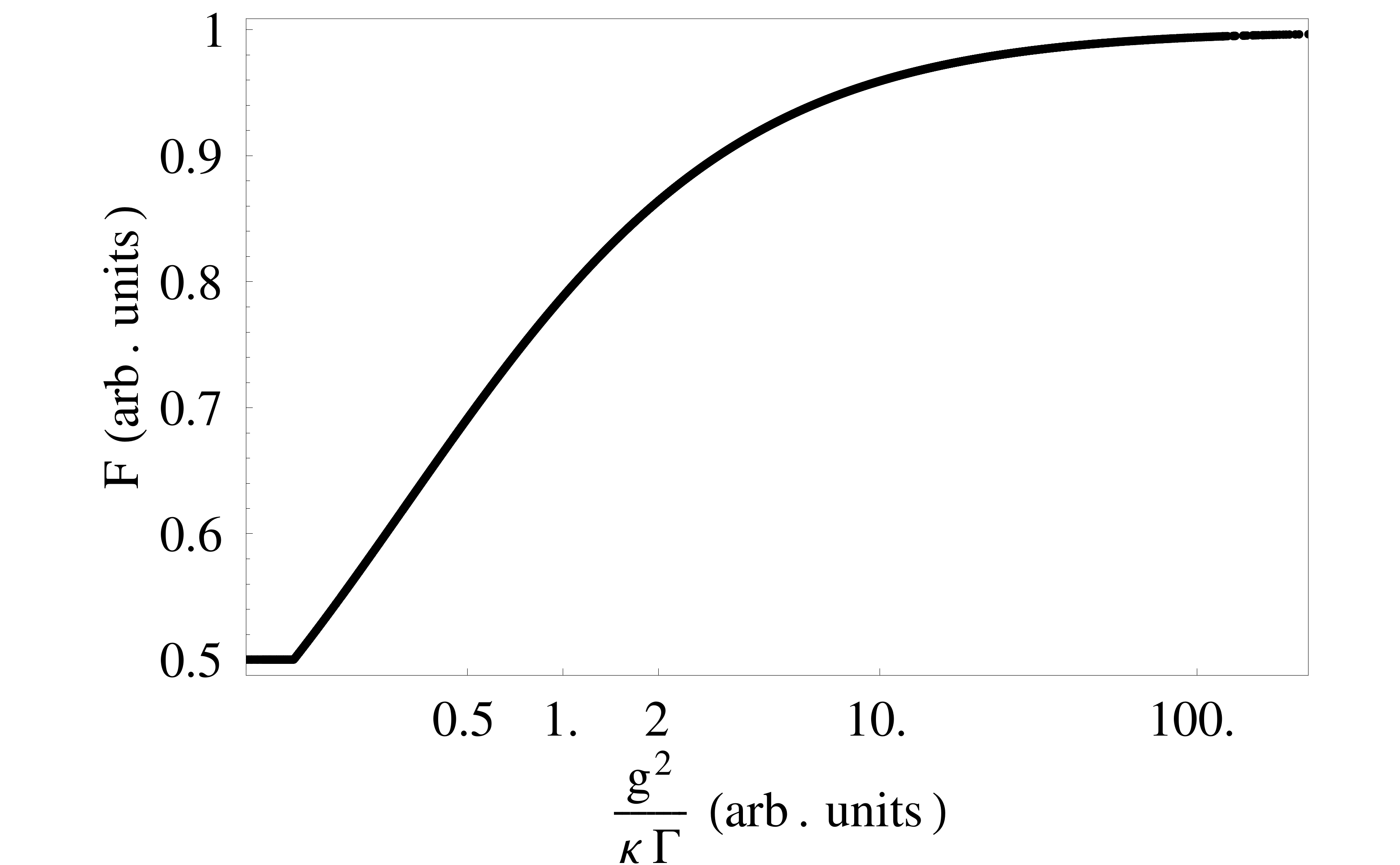}
\end{tabular}
\caption{\label{fig:fidreg} (Color online) (a) Comparison of SNR and total number of scattered photons from Sec. \ref{sec:Sch} and the fidelity for $\Delta\omega=5\;\mathrm{meV}$ versus $\alpha_{\textrm{\tiny IN}}$. (b) Maximal possible fidelity for different ratios of $\frac{g^2}{\kappa\Gamma}$ logarithmically scaled. At the border of the inequality Eq. \eqref{eq:regime}, we have $F\sim0.7$.}
\end{figure}

Equation \eqref{eq:fid2} is now evaluated and displayed in Fig. \ref{fig:fid1} for the same parameters at each subsystem, chosen to be $g=0.15\;\mathrm{meV}$, and $\kappa=0.05\;\mathrm{meV}$, which seems experimentally realistic and means a high-fidelity regime [see also Fig. \ref{fig:fidreg} (b)], while $\alpha_{\textrm{\tiny IN}}$ and $\Delta\omega$ are varied. The range for possible detunings is between $1$ and $10\;\mathrm{meV}$, as smaller detunings would lead to significant electronic excitations, whereas for larger detuning, one drives unwanted remote (light-hole) transitions and we start to lose the polarization-spin correspondence, as will be discussed in Sec. \ref{sec:Light}. A measurement window of $x_c=0.3$ leaves a success probability of $\sim25\%$ in the region of interest. A magnetic field of $B_z=1\;\mathrm{T}$ is used but this does not significantly change the results. A detuning of $\Delta\omega<2\;\mathrm{meV}$ is obviously not a good choice as the fidelity is bad and becomes also strongly $B$-field dependent.

A trade-off between the signal strength and decoherence, similar as in Ref. \onlinecite{Loo} where the situation of photon losses for far distant qubits was discussed, compared to photon losses due to scattering here, becomes obvious: At low signal, decoherence is unimportant but the uncertainty of the measurement does not allow us to project into an entangled state with high probability, whereas at higher signal intensity, decoherence via Rayleigh scattering becomes crucial. To establish a link to the estimates from Eq. \eqref{eq:estimates}, we note that the distinguishability $d_{qq}$ corresponds to the SNR and plot it together with the number of scattered photons and the resulting fidelity in Fig. \ref{fig:fidreg}(a). The criteria $SNR>1$ and $n_{scatt}<1$ used at the beginning were obviously quite good in terms of roughly revealing the regime of high fidelity. The necessary condition for the operation regime estimated in Sec. \ref{sec:Sch} to be $g^2>\frac{\kappa\Gamma}{2}$ can now be tested, as shown in Fig. \ref{fig:fidreg}(b), where we calculated the maximum possible fidelity for varying $\frac{g^2}{\kappa\Gamma}$ (with $\Gamma$ fixed at $0.002\;\mathrm{meV}$) which can be compared to the estimated condition from Eq. \eqref{eq:regime}, yielding at the boarder of the inequality $F\sim0.7$. For the parameters $g=0.15\;\mathrm{meV}$ and $\kappa=0.05\;\mathrm{meV}$ we have $\frac{g^2}{\kappa\Gamma}=225$, thus they belong to the high-fidelity $\sim0.99$ regime.

\begin{figure}[h]
\includegraphics[width=7cm]{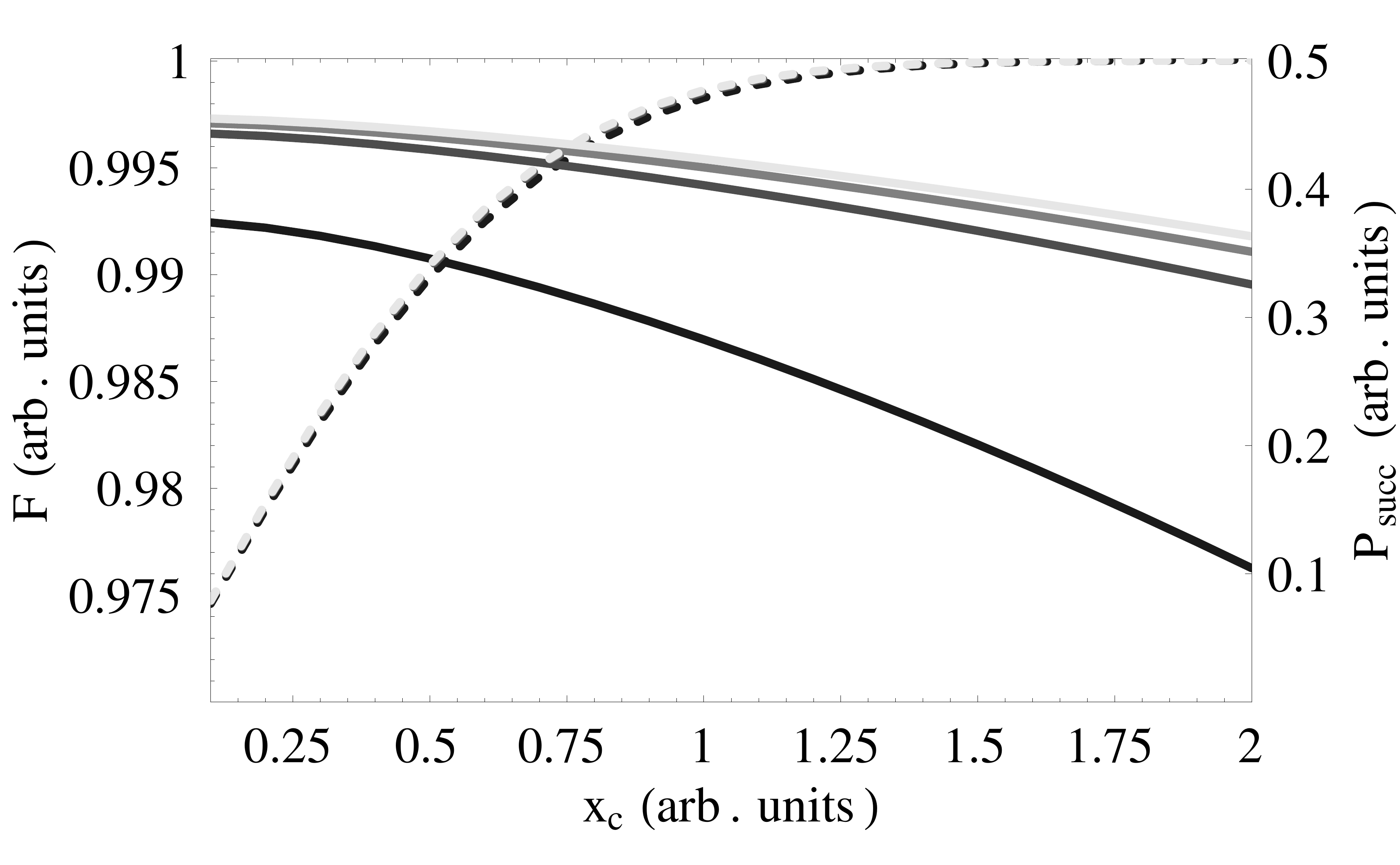}
\caption{\label{fig:PsuccDetail}Fidelity (solid lines) and success probability (broken lines) versus measurement window for various detunings of $2\;\mathrm{meV}$, $4\;\mathrm{meV}$, $6\;\mathrm{meV}$, and $10\;\mathrm{meV}$ corresponding to the gray level changing from dark to light gray. }
\end{figure}

In more detail, the connection between success probability and fidelity is displayed in Fig. \ref{fig:PsuccDetail}, showing how much fidelity we lose if we require maximal success probability. 

\subsection{Discussion of two-sided cavities\label{subsec:Dou}}

Coupling one-sided cavities requires the use of AOMs and thus it seems appealing to use two-sided cavities and directly send the transmitted light to another cavity. However, we must cope with the fact that the reflected light of a double two-sided cavity structure containing QDs carries as much information about the spin state as the transmitted light. Additionally, there might also be internal reflections between the cavities when the laser is not perfectly on resonance. A general discussion of this case can be found in Ref. \onlinecite{Poe}. 

We consider a situation with left-incident light.  The formulas derived for one-sided cavities are adapted by considering that the leaking of the light from the cavity is twice as high (assuming identical mirrors), while the coupling constant $\sqrt{\kappa}$ of the output remains the same. This decreases the number of photons in the cavity by a factor of four. The phase shift of the output in transmission is the same as that of the cavity field and thus $4$ times smaller as compared to Eq. \eqref{eq:sigOS}. If there are reflections, we also have to consider that the signal is deteriorated due to reflected light by the pulse area of the second cavity, i.e., by a factor of  $\sqrt{\Phi_B\kappa_B}$, which comes in as the normalization factor of the output [analogous to Eq. \eqref{eq:bout} for the case of one-sided cavities]. 

\begin{figure}[h]
\begin{tabular}{l}
(a)\\
\includegraphics[width=6cm]{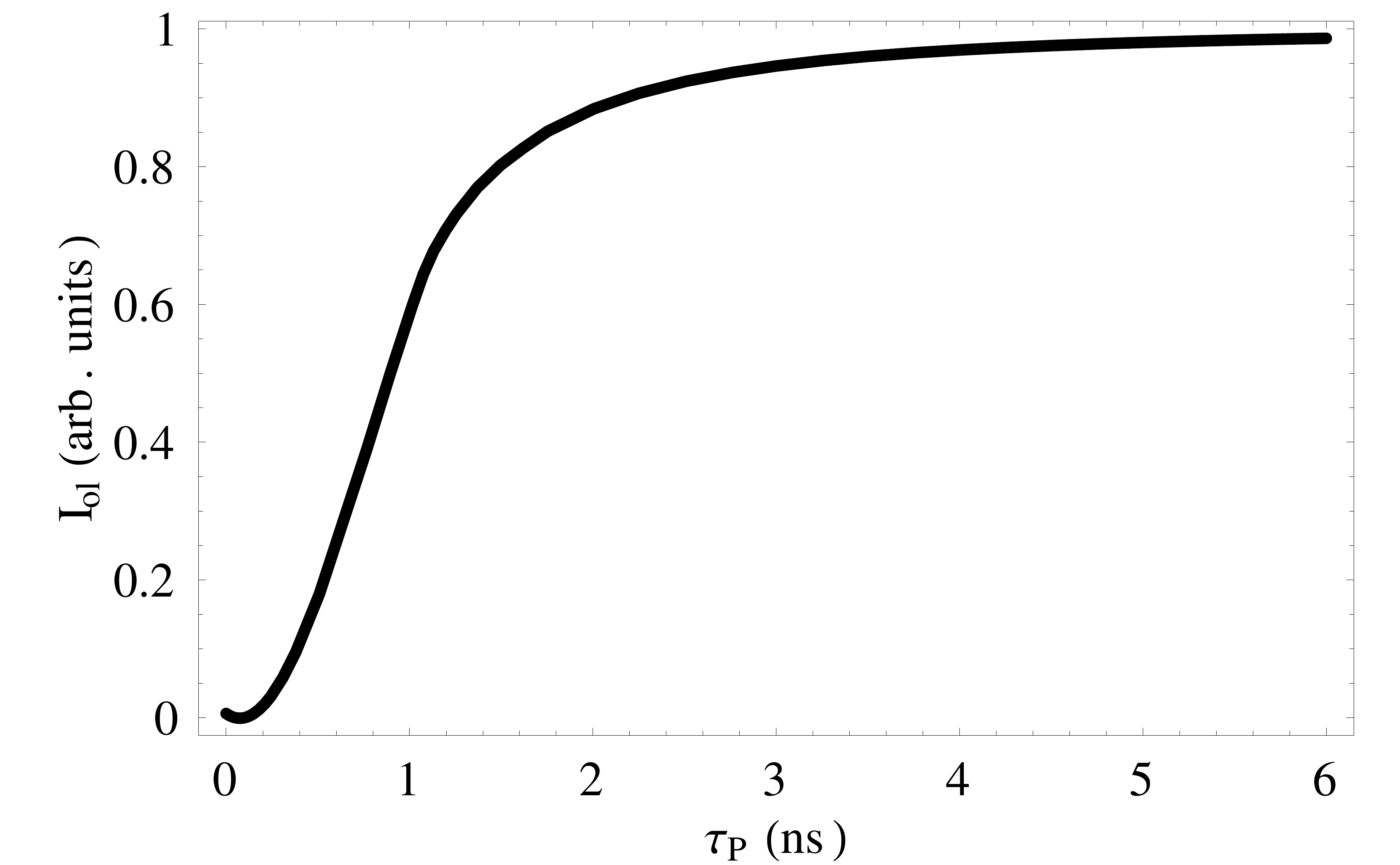}\\
(b)\\
\includegraphics[width=6cm]{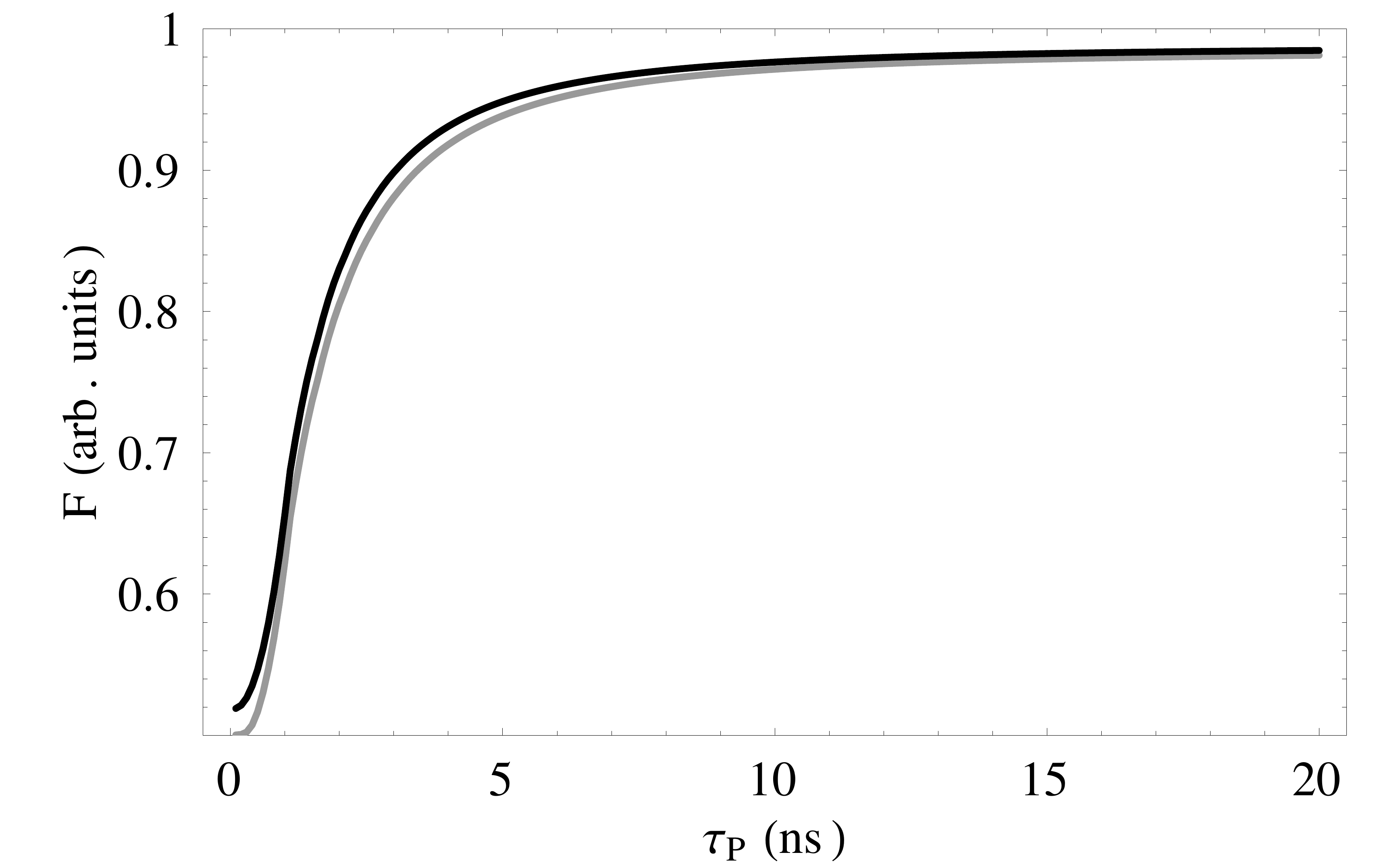}
\end{tabular}
\caption{\label{fig:OL}(a) Overlap integral $I_{ol}$ versus pulse length. (b) Maximal possible fidelity for a two-sided cavity versus pulse length $\tau_{\textrm{\tiny P}}$ for $\Delta\omega=2\;\mathrm{meV}$ (gray) and $\Delta\omega=10\;\mathrm{meV}$ (black) for $x_c=0.7$, yielding $P_{succ}>40\%$. Smaller $x_c$ do not appreciably increase the fidelity.}
\end{figure}

Entanglement generation relies upon detecting the output of the second cavity where the light carries information about both subsystems. As a two-sided cavity has outputs with spin information at any mirror, entanglement is destroyed when the reflected light can in principle be detected. Due to the boundary conditions at one cavity mirror (cf. Sec. \ref{subsec:Cav}), on resonance (with the empty cavity) reflection for light with the polarization corresponding to the active spin state and thus the sign of the $y$-polarized component depends on spin orientation (but the $x$-polarized component does not). With $\tilde{d}_{xq}:=\frac{\alpha_{\textrm{\tiny IN}}}{2}\frac{g_x^2}{\kappa \Delta\omega_{xq}}$ the $y$-polarized output of the double-cavity system at the driven (left) mirror, i.e., the  $y$-component of the reflected light, is then approximately, for the two spin configurations of interest $|g_1 g_0\ra$ and $|g_0 g_1\ra$,
\bea\label{eq:TS}
|\beta^{g_1g_0}\ra_{yL}&:=&|\tilde{d}_{A1} F_{\textrm{\tiny IN}}(t)-\tilde{d}_{B0}F_{\textrm{\tiny IN}}(t-2 t_{prop})\ra_{yL}\;,\\
|\beta^{g_0g_1}\ra_{yL}&:=&|-\tilde{d}_{A0} F_{\textrm{\tiny IN}}(t)+\tilde{d}_{B1}F_{\textrm{\tiny IN}}(t-2 t_{prop})\ra_{yL}\;,\nonu
\eea
respectively. Thus, for equal QDs, we observe two subsequent pulses of opposite amplitude, the order of which depends on the spin states. Tracing over this degree of freedom leads to additional decay of the fidelity due to the decay of $\la \beta^{g_1,g_0}|_{yL}|\beta^{g_0,g_1}\ra_{yL}$ with details given in Appendix \ref{app:A}. If the pulse length is short enough such that the reflected pulses do not overlap, the effect has its maximum and the fidelity practically decays completely. For longer pulse lengths, the overlap integral $I_{ol}(\tau_{\textrm{\tiny P}}):=\int F_{\textrm{\tiny IN}}(t)F_{\textrm{\tiny IN}}(t-2t_{prop}) dt$ approaches unity [see Fig. \ref{fig:OL}(a)]. The dependence of the fidelity on the pulse length is shown in Fig. \ref{fig:OL}(b) for small ($2\;\mathrm{meV}$) and large ($10\;\mathrm{meV}$) detunings with a success probability $>40\%$. \footnote{Photon number detection of the reflected light of each cavity would preserve the fidelity, even for shorter pulse lengths.}

\section{Nonidentical Quantum Dots\label{sec:Non}}

In experiments, the ideal results from the last section will not apply since it is not very likely to find two self-assembled QDs of same frequency and same $g$. Here, we show how to overcome this problem by several different strategies depending on the difference in trion transition energies 
$\nu_A$ and $\nu_B$ of the two quantum dots.

\begin{figure*}[t]
\begin{tabular}{l l}
(a)&(b)\\
\includegraphics[width=7cm]{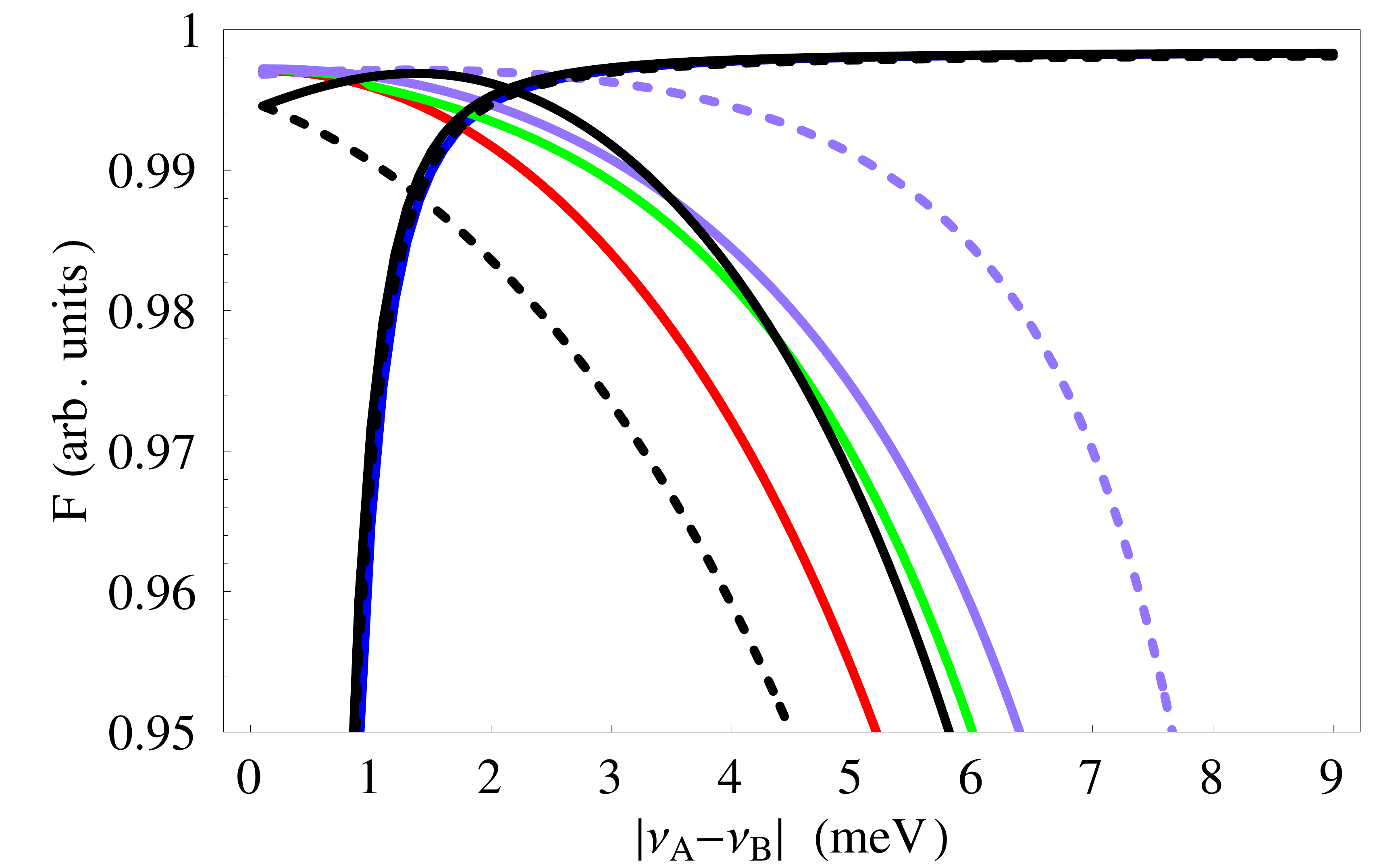}&\includegraphics[width=7cm]{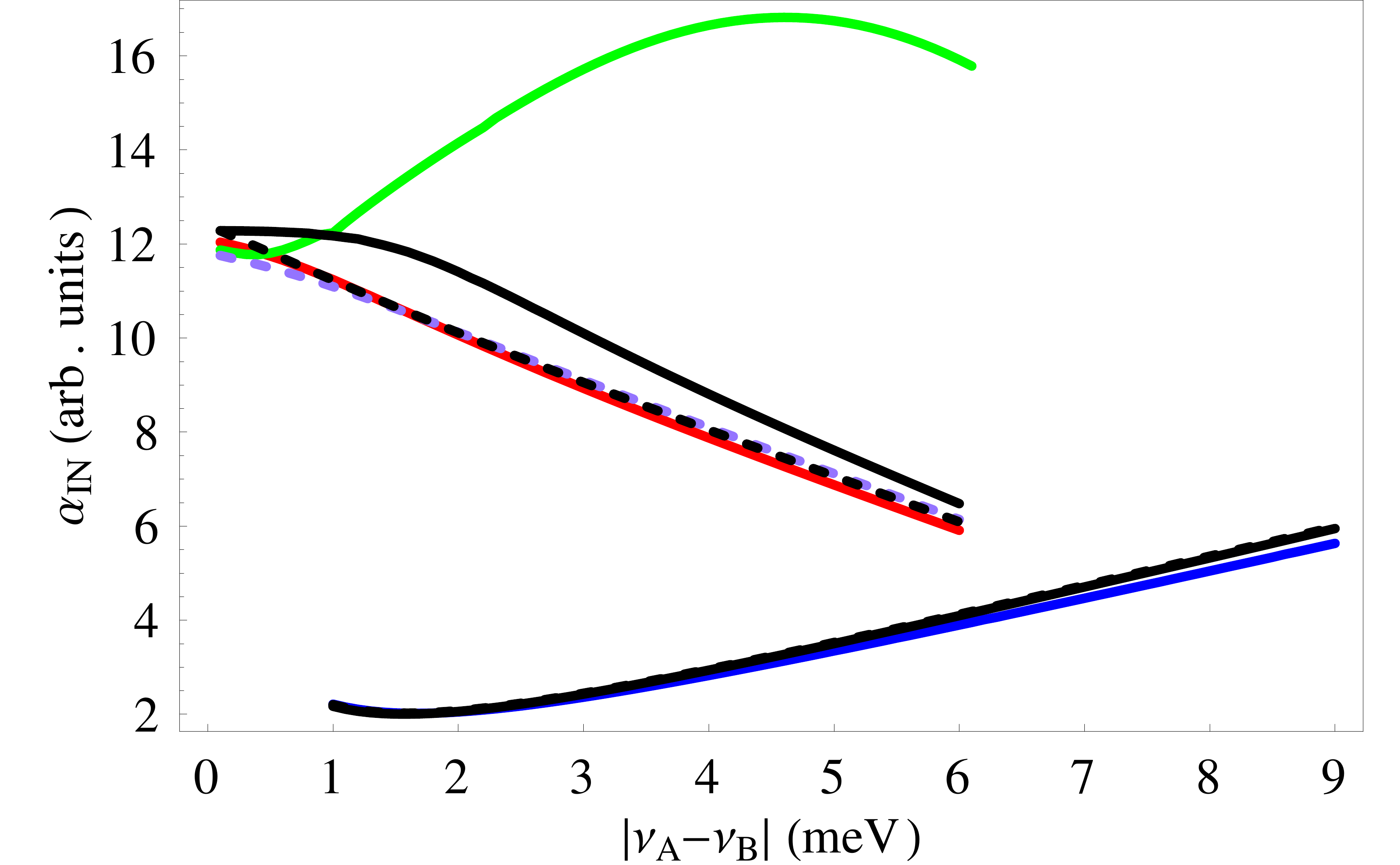}\\
(c)&(d)\\
\includegraphics[width=7cm]{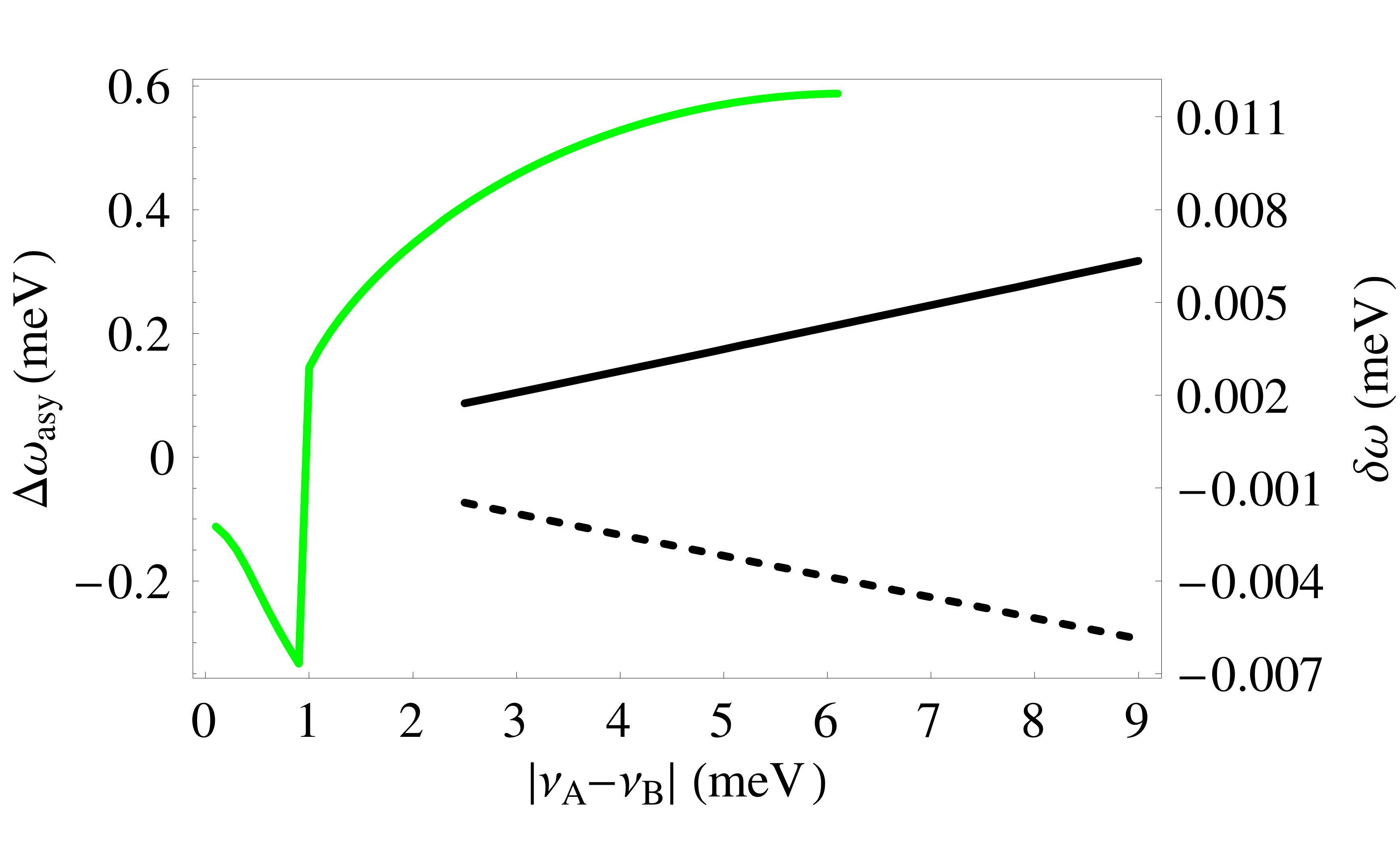}&\includegraphics[width=7cm]{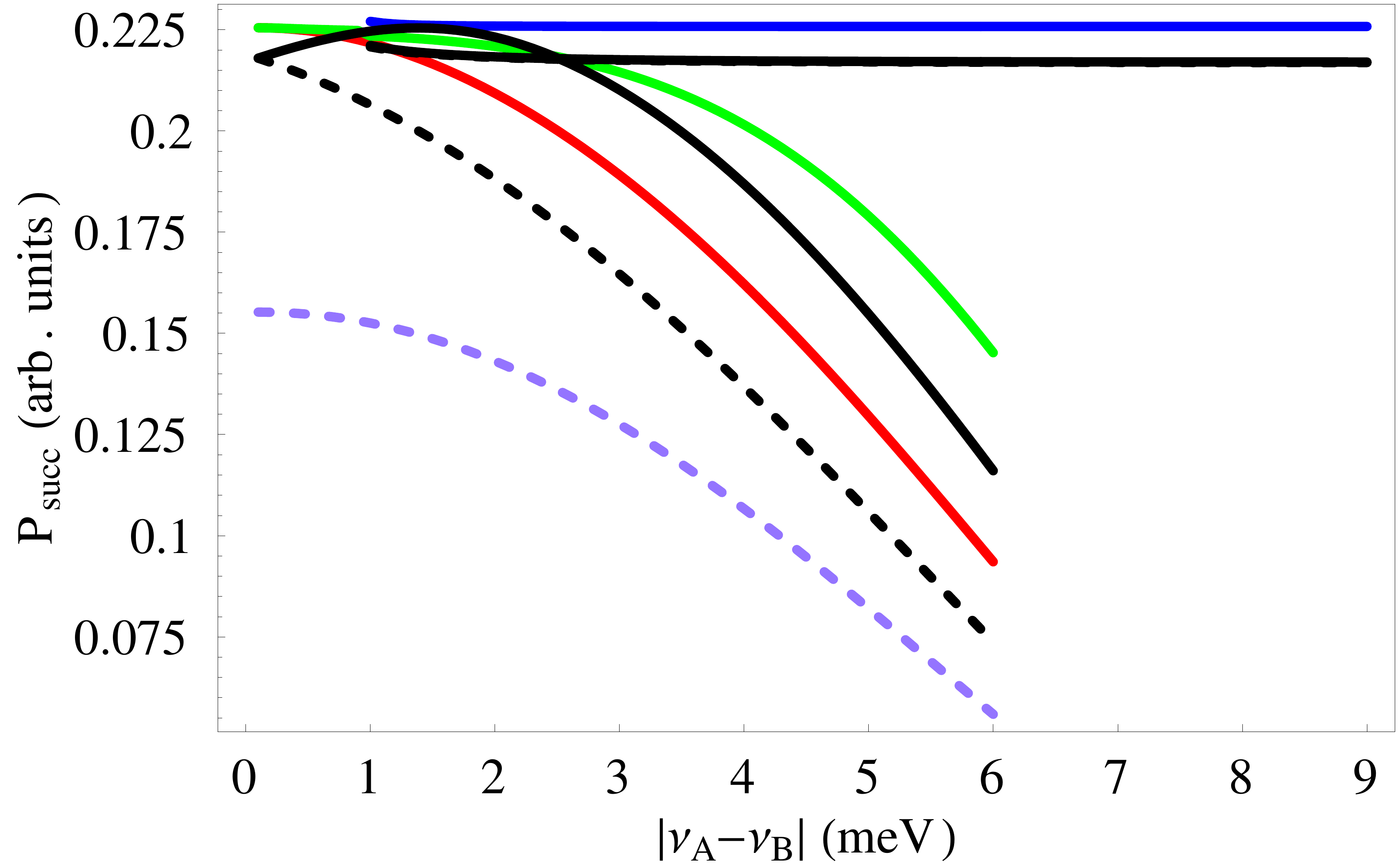}
\end{tabular}
\caption{\label{fig:Diff1} (Color online) (a) Outline of the fidelity versus $|\nu_A-\nu_B|$ for different strategies. Tuning in between the two QD resonances does not work for small $|\nu_A-\nu_B|$ (blue line), where redshifting is more appropriate (red line) with maximal detuning being $10\;\mathrm{meV}$. One could also detune the cavity containing the QD with larger frequency (smaller detuning when redshifting) slightly from the laser (green line) or chose a smaller measurement window of $x_c=0.2$ (violet line). Then, one obtains a smaller success probability as compared to $x_c=0.3$, which was the choice for all other lines. The black lines are for $g_A=0.14\;\mathrm{meV}$ with $g_B=0.15\;\mathrm{meV}$ (solid black line) and $g_B=0.14\;\mathrm{meV}$ with $g_A=0.15\;\mathrm{meV}$ (broken black line), respectively. The second case deviates a lot from the first for red shifting and we should accept a lower success probability ($x_c=0.1$) yielding better fidelity (broken violet line), whereas for tuning in between, the lines are indistinguishable. In (b), the optimal photon number is shown which is of course bigger when the laser is detuned from the cavity (green line). When $g_A\neq g_B$, the laser is not tuned in symmetrically between the QD transition energies but asymmetrically with $\Delta\omega_{asy}=\frac{|\nu_B-\omega_L|-|\nu_A-\omega_L|}{2}$, as shown in (c) (black lines), together with the optimized cavity-laser detuning $\delta\omega=|\omega_0-\omega_L|$ at one cavity (green line). All the lines are interpolations of the evaluated points. In (d), we plot the success probability. }
\end{figure*}

For large detuning of the QDs $\mid\nu_A-\nu_B\mid$, one can tune the laser symmetrically in between the QD resonances instead of redshifting in order to balance the detuning for the two subsystems. Then, one produces the Bell triplet state $|\phi^+\ra=\frac{1}{\sqrt{2}}(|00\ra+|11\ra)$ instead of the singlet state since the sign of the phase shift depends on the sign of the detuning. Thus, the phase shift of the first cavity differs by a sign from that of the second cavity and the role of $|01\ra,\;|10\ra$ and $|00\ra,\;|11\ra$ is reversed (see also Appendix \ref{app:B}). With the help of the formalism introduced in the preceding sections with the simple analytic formula for the fidelity, it is now easy to analyse the dependence of the fidelity on varying $\Delta\omega_A$ and $\Delta\omega_B$ by finding the optimal parameters numerically, as shown in Fig. \ref{fig:Diff1}. The resulting fidelity becomes bad when $|\nu_A-\nu_B|<2\;\mathrm{meV}$ due to Rayleigh scattering, as shown in Fig. \ref{fig:Diff1} (blue line).

In order to have good fidelity when the QDs differ not too much, it is advantageous to redshift the cavity frequency such that it is detuned from the QD with the lower transition frequency by the maximal allowed value, which we estimated in Sec. \ref{subsec:Res} to be about $\Delta\omega=10\;\mathrm{meV}$. In this case, the relative difference of the phase shifts at each QD is minimized, while light-hole mixing effects remain small. By optimizing the photon number, we get high fidelity for $|\nu_A-\nu_B|< 2\;\mathrm{meV}$, as shown by the red line in Fig. \ref{fig:Diff1}.

For small detunings it may also pay to slightly detune the cavity with the "better" (higher transition energy) QD from the laser such that the phase shift at this QD becomes of the same size as that one at the other, "worse" QD (see Appendix \ref{app:B} for details). We find now optimal $\alpha_{\textrm{\tiny IN}}$ and laser-cavity detuning at one of the subsystems (here $A$) $\delta\omega=\omega_L-\omega_{0A}$ by optimizing the fidelity (green line). However, by doing this, the signal which distinguishes the entangled state from product states $|d_{10}\ra_y$ and $|d_{01}\ra_y$  (cf. Sec. \ref{sec:Fid}), which is the sum of both phase shifts at either cavity, decreases which lowers the SNR. Thus, this method does not work for arbitrary $\delta\omega$. The same effect could be achieved by decreasing the measurement window $x_c$ from $0.3$ to $0.2$, which, however, decreases the success probability (violet line).

Realistically, we have also different QD-cavity coupling constants $g_A$ and $g_B$ and we can also compensate for this by either tuning in between for big $|\nu_A-\nu_B|$ or redshifting otherwise if $g_A<g_B$ and $\nu_A<\nu_B$.  Thereby, we compensate for, e.g., a smaller $g_A$ with a smaller $\Delta\omega_A$ (black lines). If $g_A>g_B$ and $\nu_A<\nu_B$ and thus the QDs are even more nonidentical, tuning in between works well, but for red shifting a smaller measurement window $x_c=1$ must be chosen.

\begin{figure*}[t]
\begin{tabular}{l l}
(a)&(b)\\
\includegraphics[width=7cm]{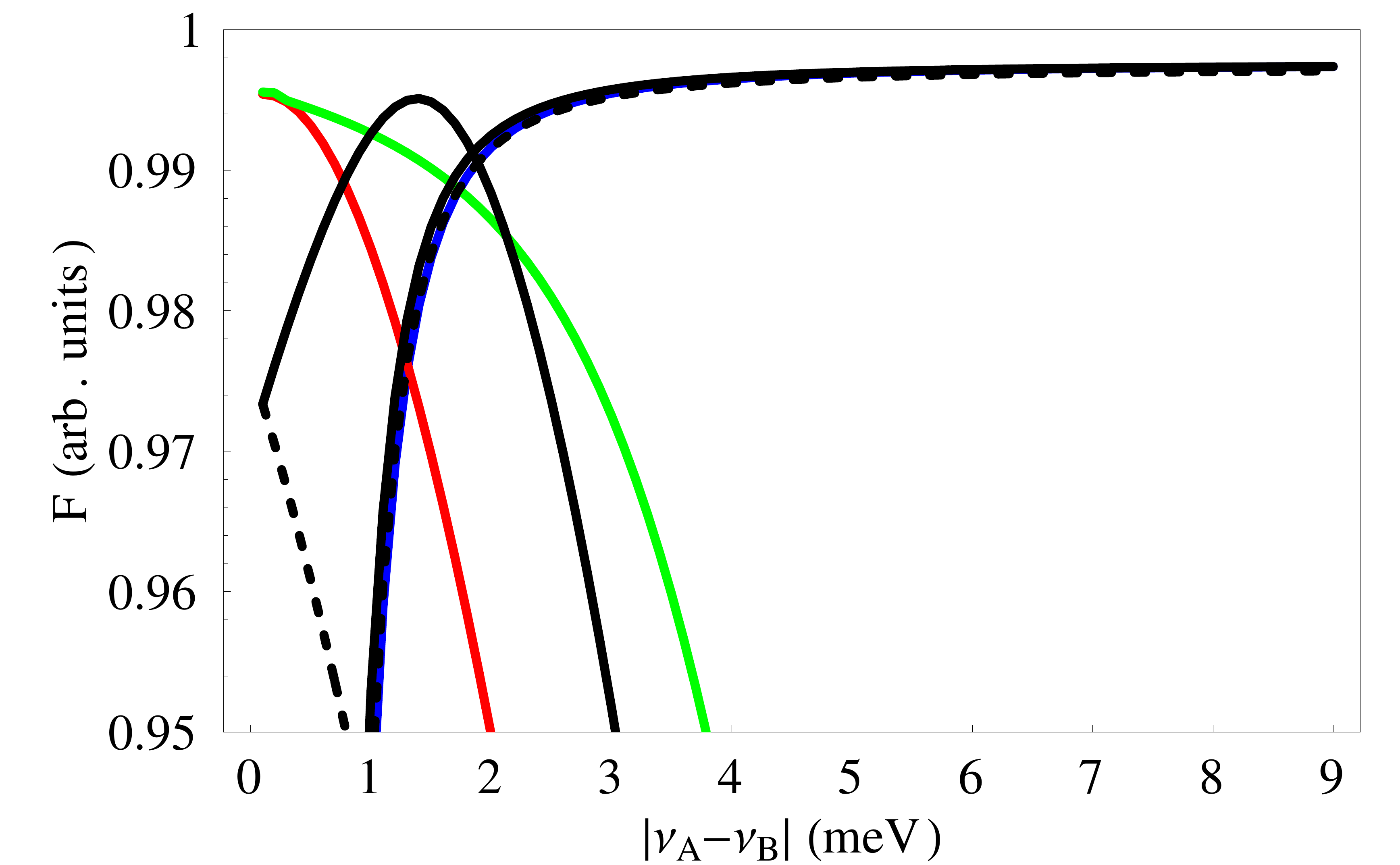}&\includegraphics[width=7cm]{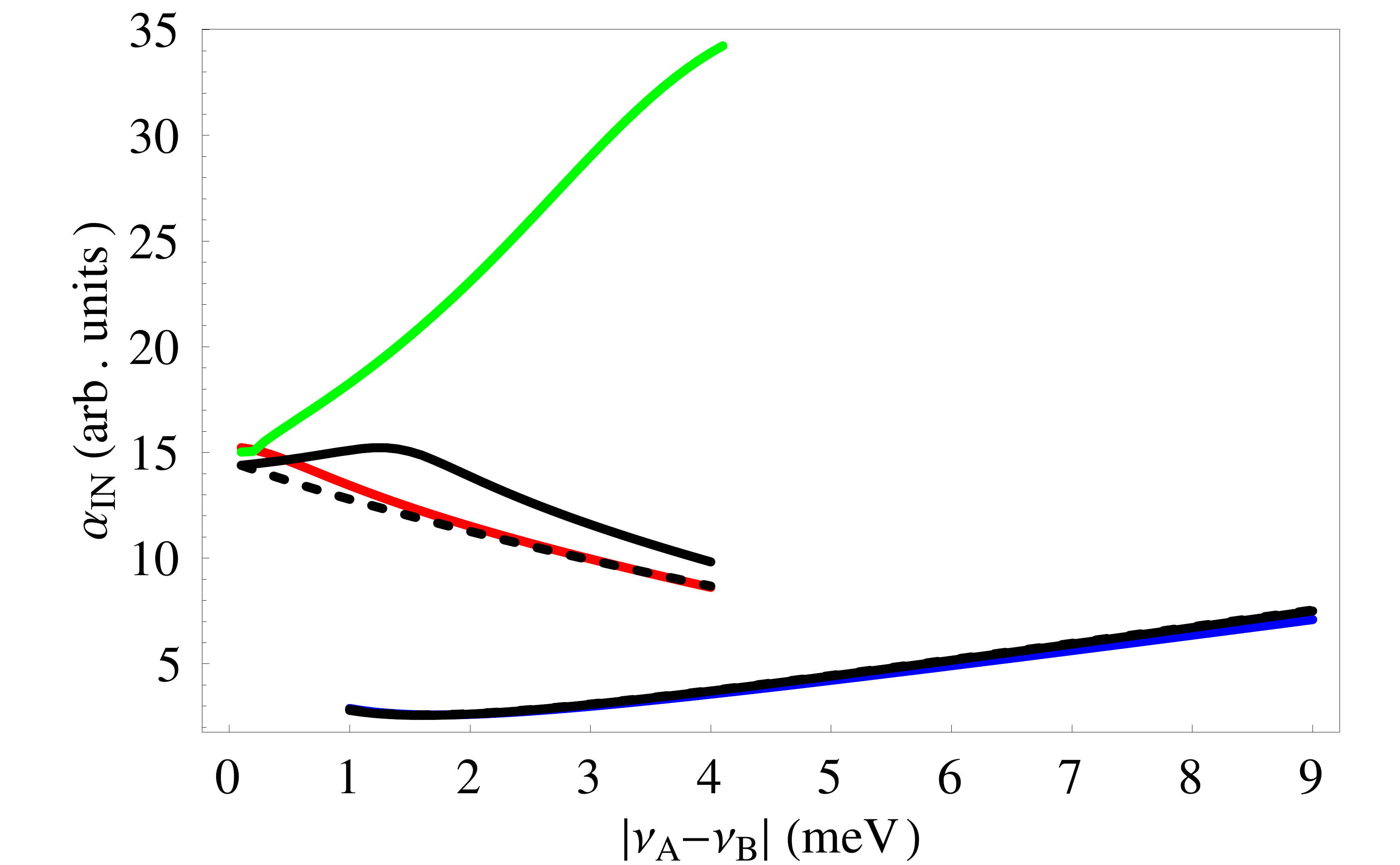}\\
(c)&(d)\\
\includegraphics[width=7cm]{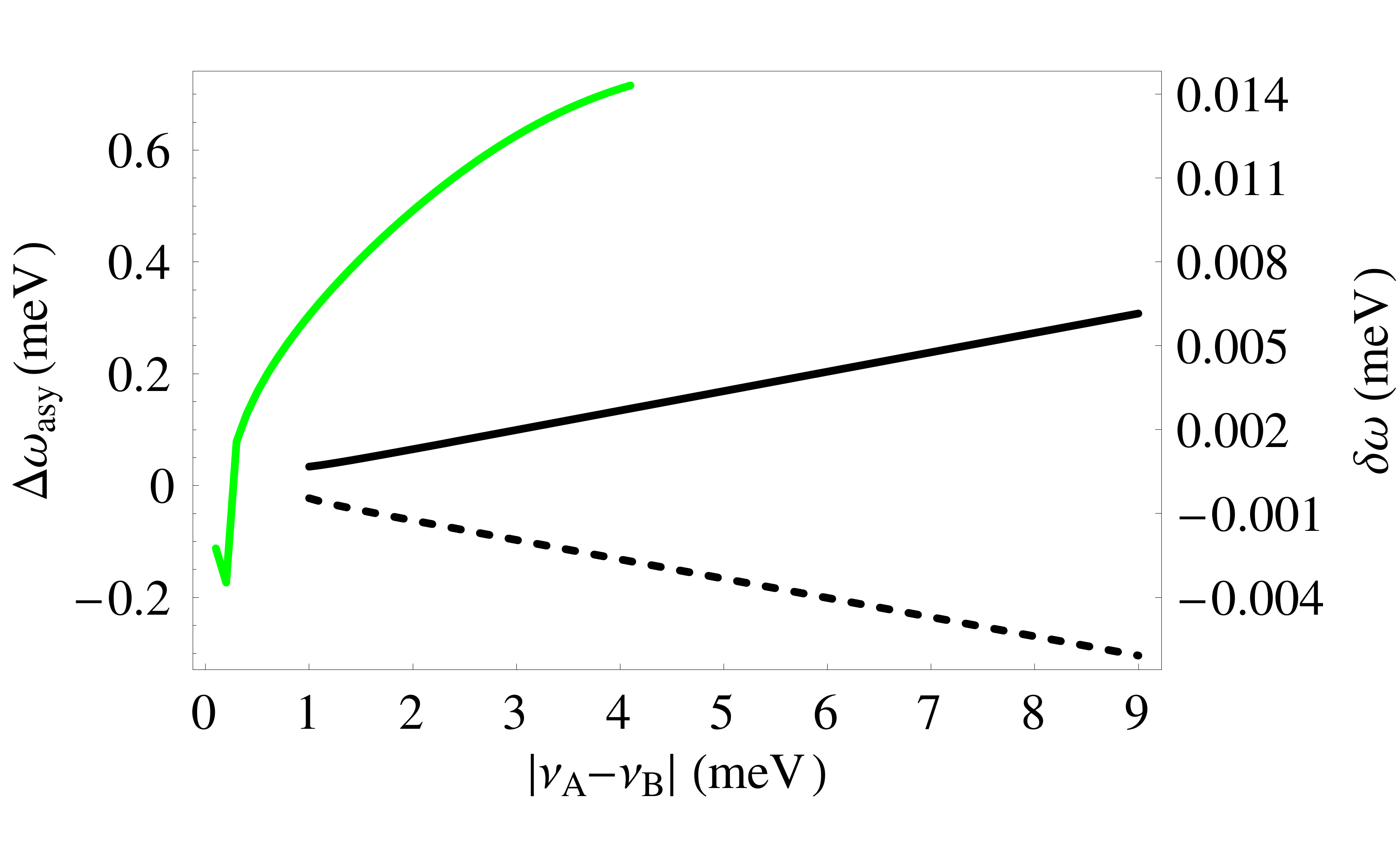}&\includegraphics[width=7cm]{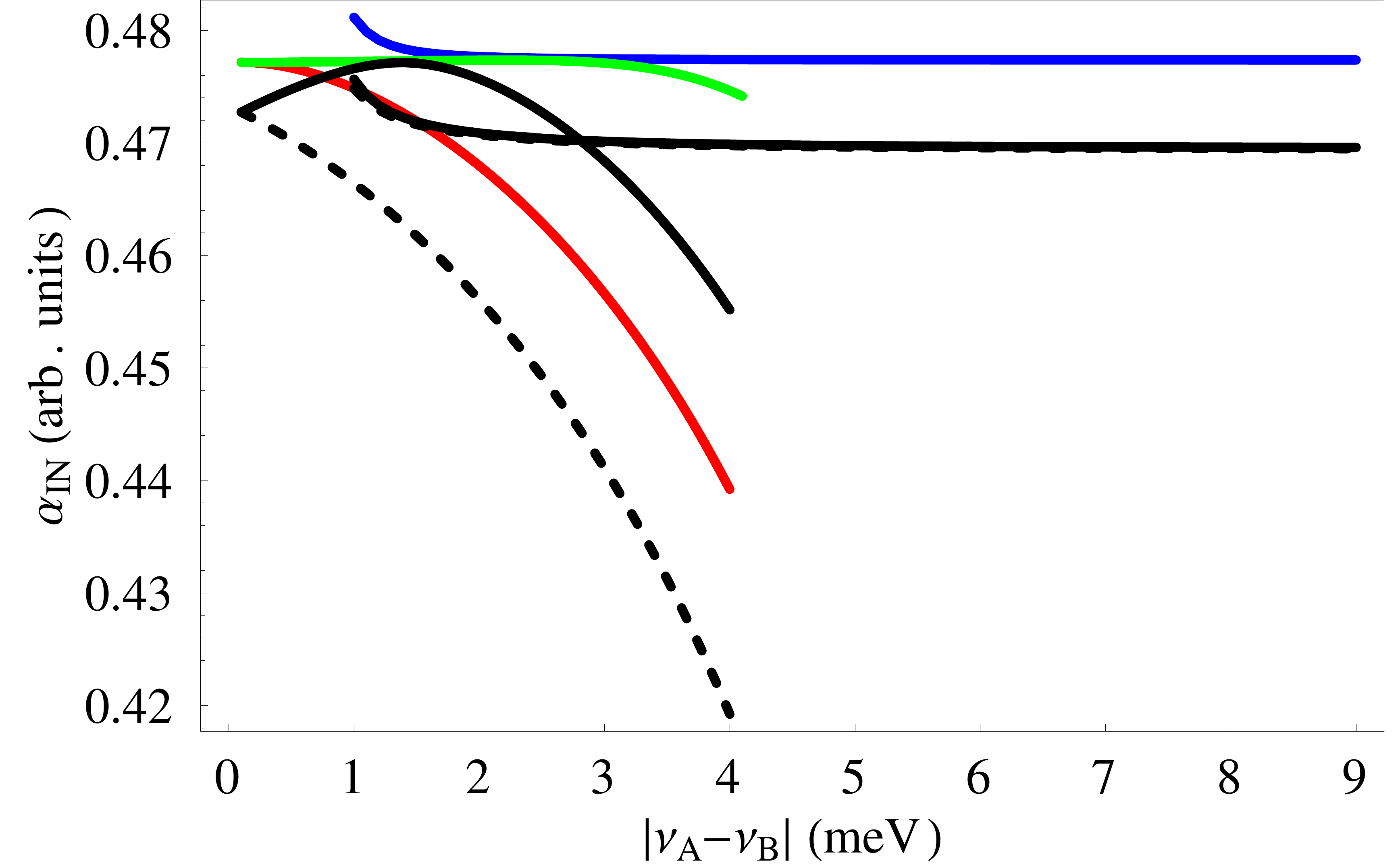}
\end{tabular}
\caption{\label{fig:Diff1xc15} (Color online) (a) Outline of the fidelity versus $\Delta=|\nu_A-\nu_B|$ for different strategies with same encoding as in Fig. \ref{fig:Diff1}. A larger measurement window of $x_c=1$ was chosen, such that the success probability in (d) increases significantly, with values above $P_{succ}=0.47$. In (b) and (c), the calculated optimal photon numbers and detunings, respectively, are shown. }

\end{figure*}

In order to have higher success probability, we choose a larger measurement window of $x_c=1$ and examine the same strategies as before in Fig. \ref{fig:Diff1xc15}. While the success probability can now be increased above $P_{succ}=0.47$, simple redshifting (red line) is now less suited to compensate for different QDs. Instead, it pays now to detune one cavity slightly from the laser (green line). Tuning in between for different QD transition frequencies or different light-matter coupling constants still works well.

\section{Light-hole transitions\label{sec:Light}}
\begin{figure}[t]
\includegraphics[width=7cm]{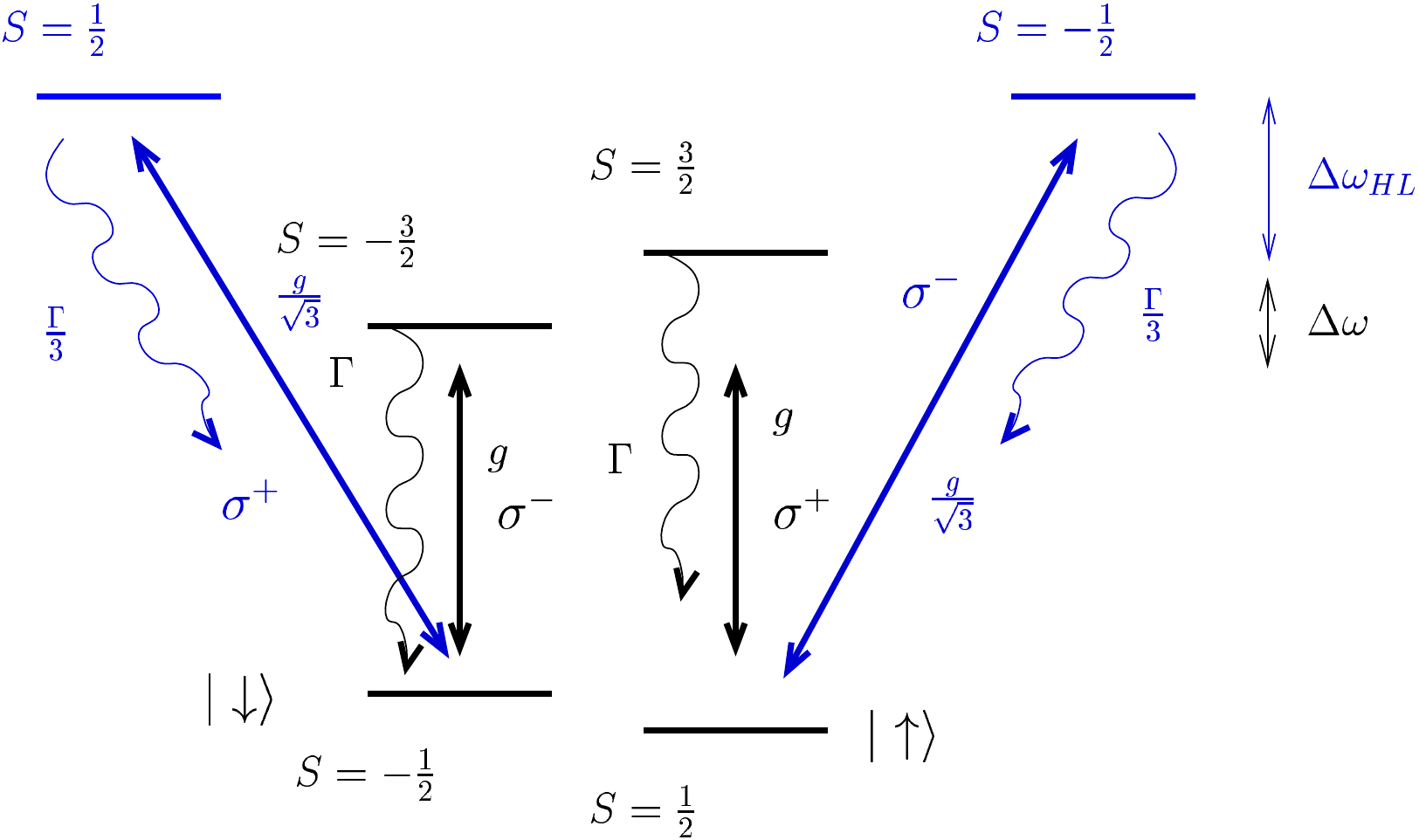}
\caption{ (Color online) Level scheme including the light holes which have different spins $S$ than the heavy holes. Each spin ground state is now addressed by light of both circular polarizations but with different coupling strengths.\label{fig:light}}
\end{figure}

So far, we considered only the dominant heavy hole (trion) transitions which is certainly justified for small detuning. Going to higher detuning or when considering nonidentical QDs and tuning in between, as in Sec. \ref{sec:Non}, requires taking into account the valence band structure, i.e., also the light-hole transitions. Effects from other, remote transitions are discussed in Ref. \onlinecite{Poe} and were found to be small for a detuning of less than 10 meV. The light-hole-associated energy levels are separated by at least $\Delta\omega_{HL}=10\;\mathrm{meV}$ from the top heavy-hole level. Moreover, their dipole matrix elements are reduced by a factor of $\frac{1}{\sqrt{3}}$ with respect to to heavy holes, see, e.g., Ref. \onlinecite{Zut}, leading to $g\rightarrow\frac{g}{\sqrt{3}}$ and $\Gamma\rightarrow\frac{\Gamma}{3}$. The allowed transitions lead to a coupling of each spin to both circular polarizations, as depicted in Fig. \ref{fig:light}. 

\begin{figure}
(a)\\
\includegraphics[width=5cm]{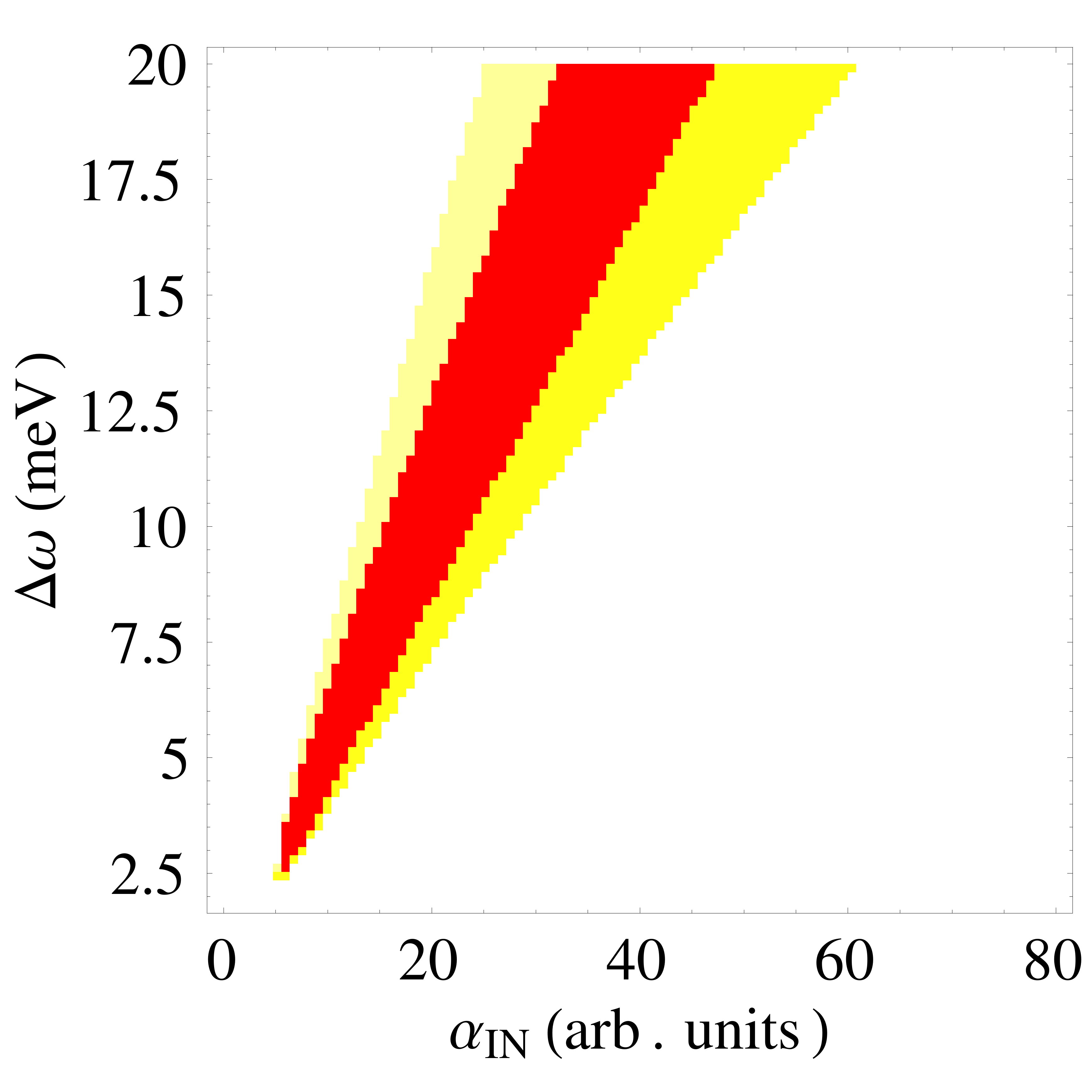}\\
(b)\\
\includegraphics[width=5cm]{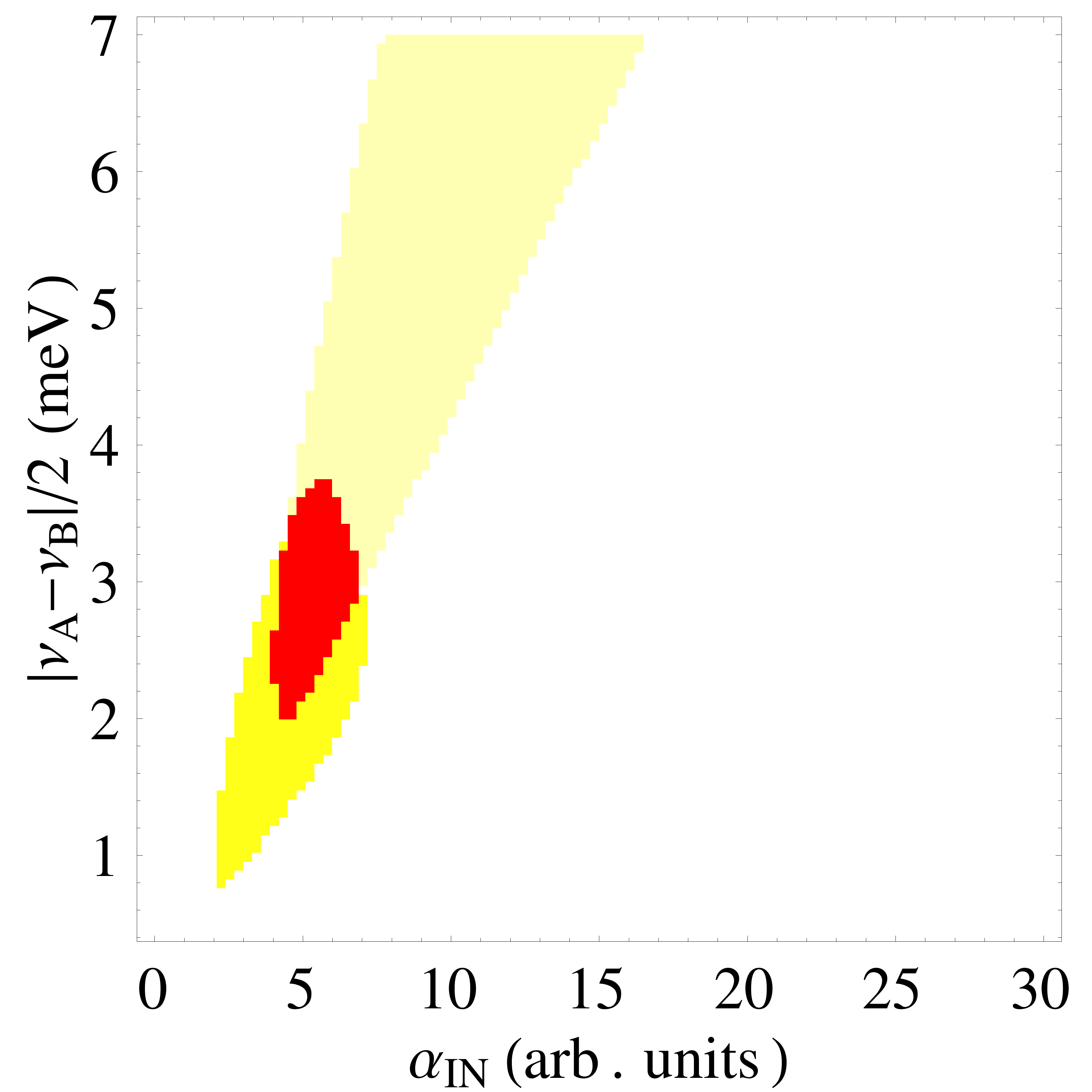}
\caption{\label{fig:light1} (Color online) The colored regions signify where $F>0.99$. The red region indicates the overlap of the case with light holes included ($\Delta\omega_{HL}=10\;\mathrm{meV}$) and no light holes. The light yellow region [left yellow region in (a) and in (b)] is only for the latter case and the darker yellow region for the first case. In (a), both QDs are redshifted equally and $P_{succ}>0.47$ ($x_c=1$), whereas in (b), the laser is tuned in between the two QDs and thus one of them is blueshifted, restricting the region of high fidelity and $P_{succ}>0.35$ ($x_c=0.6$). }
\end{figure}

\begin{figure}
\includegraphics[width=5cm]{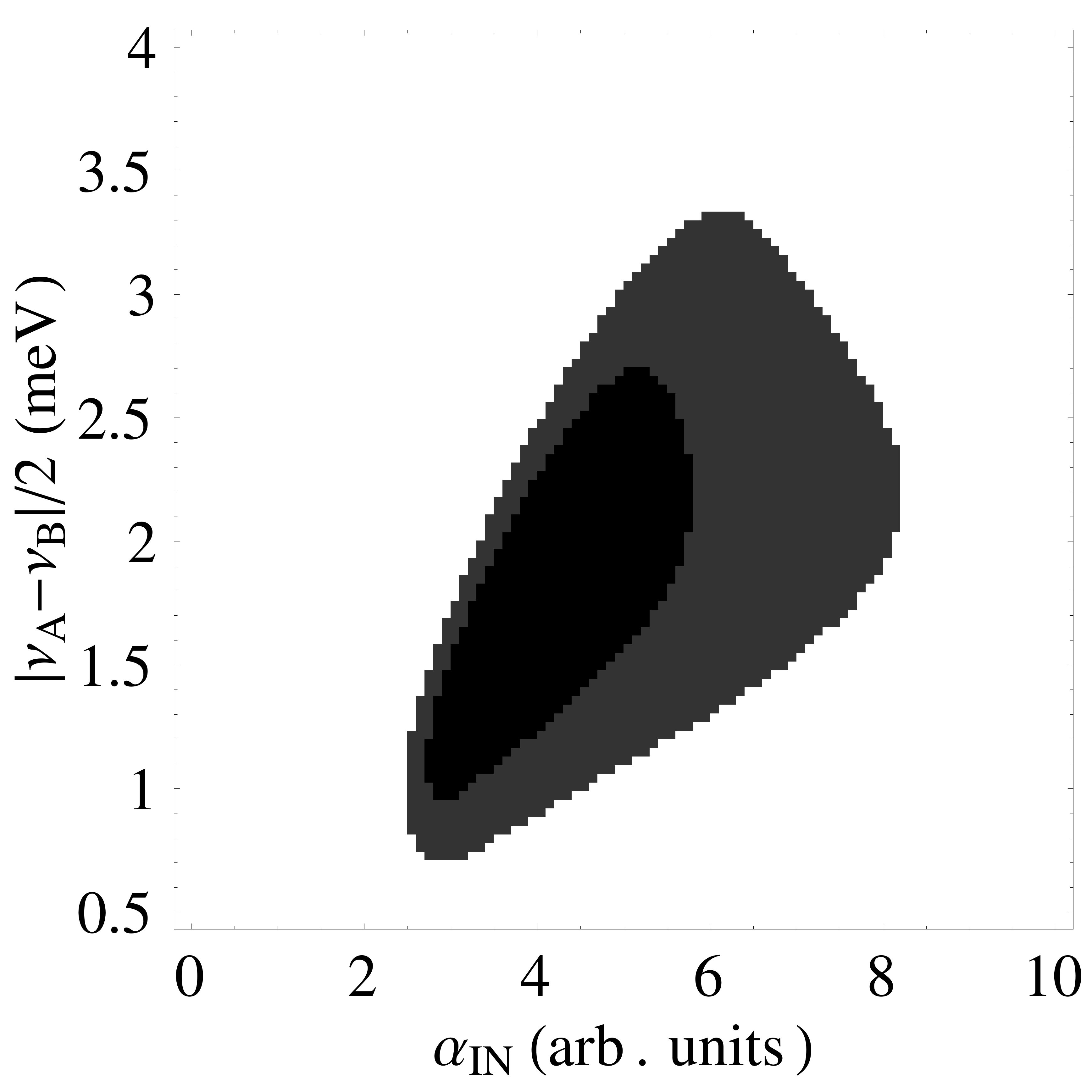}
\caption{\label{fig:light2}The black region signifies where $F>0.99$ and the gray region $F>0.98$ for the overlap region of $\Delta\omega_{HL}=10\;\mathrm{meV}$ and $\Delta\omega_{HL}=20\;\mathrm{meV}$ and tuning in between the two QDs. The success probability is $>0.49$ ($x_c=1.3$).}
\end{figure}

The Faraday rotation is thus, for a fixed spin state, reduced as now both circular polarizations acquire a finite but different phase, such that the relative phase decreases. The Stark shift becomes reduced as
\be
\frac{g^2}{\Delta\omega}\rightarrow\frac{g^2}{\Delta\omega}\Bigl(1-\frac{1}{3}\frac{\Delta\omega}{\Delta\tilde{\omega}}\Bigr)\;.
\ee
Applying a Schrieffer-Wolff transformation to the Hamiltonian including the light holes, one gets analogously to Sec. \ref{sec:Fid} Rayleigh scattering contributions for all transitions involved, but we now have two scattering contributions for each polarization. This corresponds to choosing the same reservoirs for the transitions which are coupled by the same polarization. The Lindblad operator \cite{Bre} for a certain circular polarization then reads
\be
\hat{L}_q=\sqrt{\Gamma}\bigl( \frac{g}{\Delta\omega}|g_q\ra\la g_q|+\frac{g}{3\Delta\tilde{\omega}}|g_{q'}\ra\la g_{q'}|\bigr)\ah_q\;,
\ee
where $\Delta\tilde{\omega}=\Delta\omega_{HL}\pm\Delta\omega$ for either redshifting or blueshifting, respectively. This implies that scattering of circular polarized light occurs for both spin states but at different rates and thus the scattered photons carry less information about the spin state as compared to the case without light holes. The actual rate at which coherences decay [cf. Eq. \eqref{eq:decent}] is decreased and given by the replacement
\be
\Gamma_q^R\rightarrow\Gamma_q^R\Bigl(1-\frac{1}{3}\frac{\Delta\omega}{\Delta\tilde{\omega}}\Bigr)^2\;.
\ee
For redshifting, the corrections due to light hole transitions are rather small for detunings $\Delta\omega<10\;\mathrm{meV}$ and there is still always a region of high fidelity for $\Delta\omega>10\;\mathrm{meV}$ [see Fig. \ref{fig:light1}(a)]. 
The red region signifies where the fidelity is higher than $0.99$, while $P_{succ}>0.47$ for both the cases of light holes split by $\Delta\omega_{HL}=10\;\mathrm{meV}$ and no light holes. The light yellow region signifies where this is true only for the latter case, the darker yellow region only for the case with light holes included. Thus, for redshifting, we do not have to worry about light holes also for bigger detunings. They become more crucial when considering nonidentical QDs and tuning in between the transition frequencies as in Sec. \ref{sec:Non} because then one QD is blueshifted with respect to the laser and is thus close to the light holes. The regime of high fidelity exists only for small detuning and photon number, as can be seen in Fig. \ref{fig:light1}(b) for $P_{succ}>0.35$. Although we could find regimes of high fidelity also when light holes are included, it remains the problem that we do not exactly know how much they are energetically split from heavy holes. Thus, we search for the overlap of the results for a pessimistic estimate, $\Delta\omega_{HL}=10\;\mathrm{meV}$, and a rather optimistic one, $\Delta\omega_{HL}=20\;\mathrm{meV}$. In Fig. \ref{fig:light2} we plot the region of high fidelity $>0.99$ with a high success probability of $P_{succ}>0.49$. Thus, we conclude that a fairly high fidelity together with a good success probability can be obtained also for the case of different quantum dots and the presence of light holes. Experimentally, it should be feasible to find two QDs which differ between $2\;\mathrm{meV} $ and $6\;\mathrm{meV}$ and tune in between. The discussions of this section are very convincing that QDs represent excellent systems for distant entanglement generation.


\section{Semiclassical simulation\label{sec:Sem}}

\begin{figure}[h]
\begin{sideways}
\includegraphics[width=4.2cm]{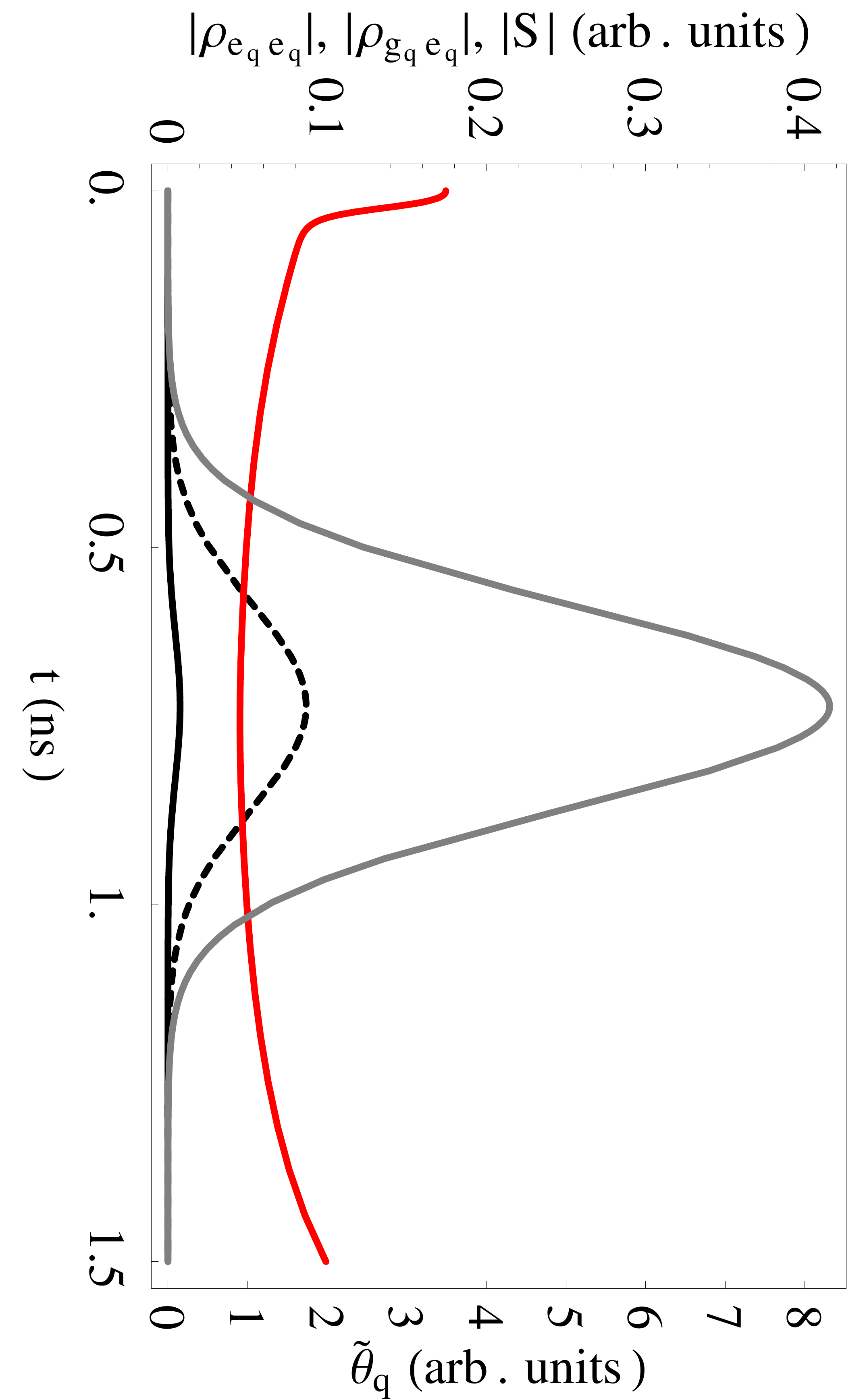}
\end{sideways}
		\caption{(Color online) The red line displays a solution of Eq. \eqref{eq:OBE1} for the phase shift $\tilde{\theta}_q(t)$ defined in Eq. \eqref{eq:sigOS}. The solid black line corresponds to $\rho_{e_q e_q}(t)$ from Eq. \eqref{eq:OBE2} showing that the population of the excited state is low, whereas the dashed line corresponds to  $\rho_{g_q e_q}(t)$. These amplitudes adiabatically follow the cavity field (gray line). Parameters as in text. 
		\label{fig:rho1}}
\end{figure}

\begin{figure*}[t]
\begin{tabular}{l l}
(a)&(b)\\
\includegraphics[width=7cm]{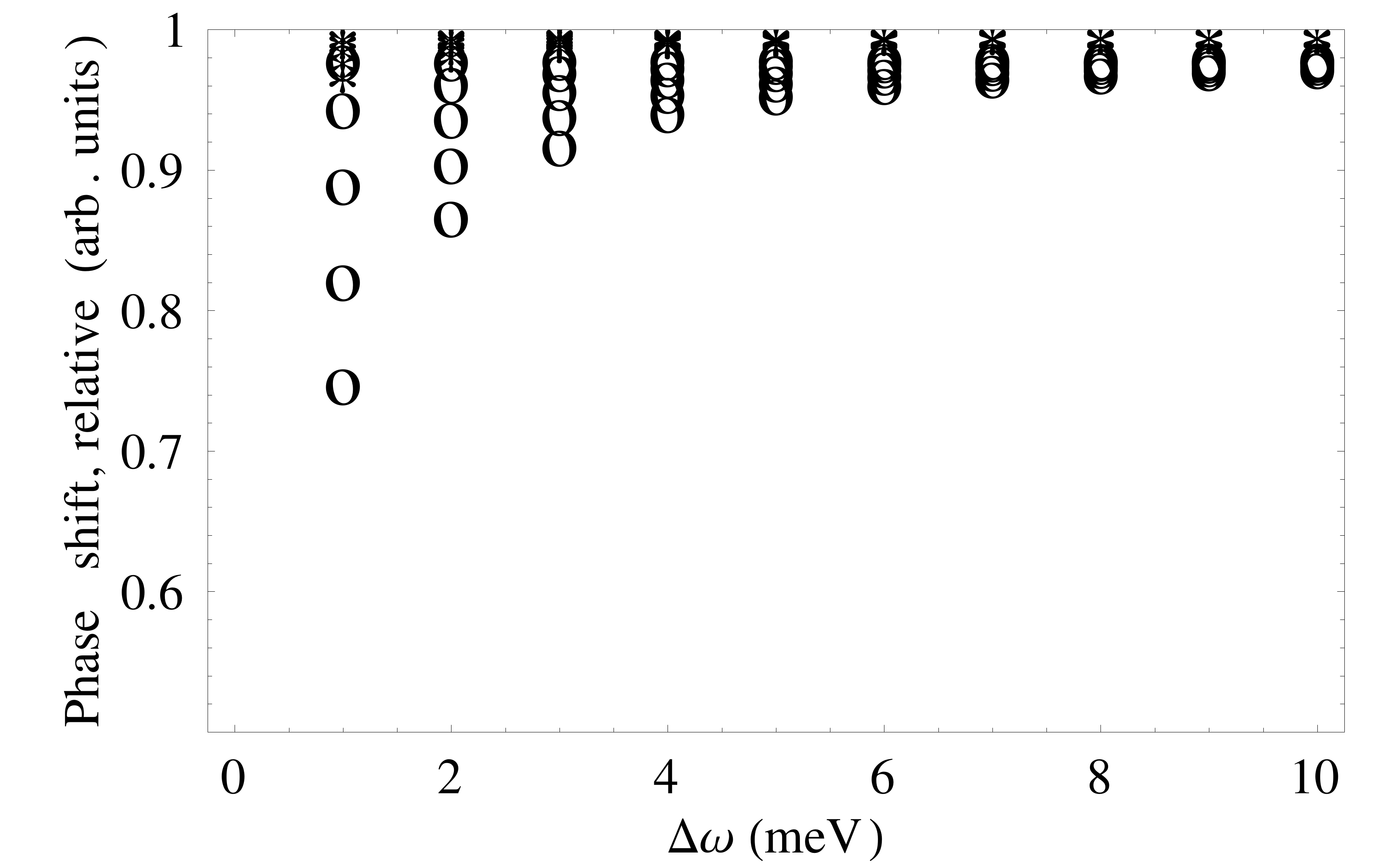}&\includegraphics[width=7cm]{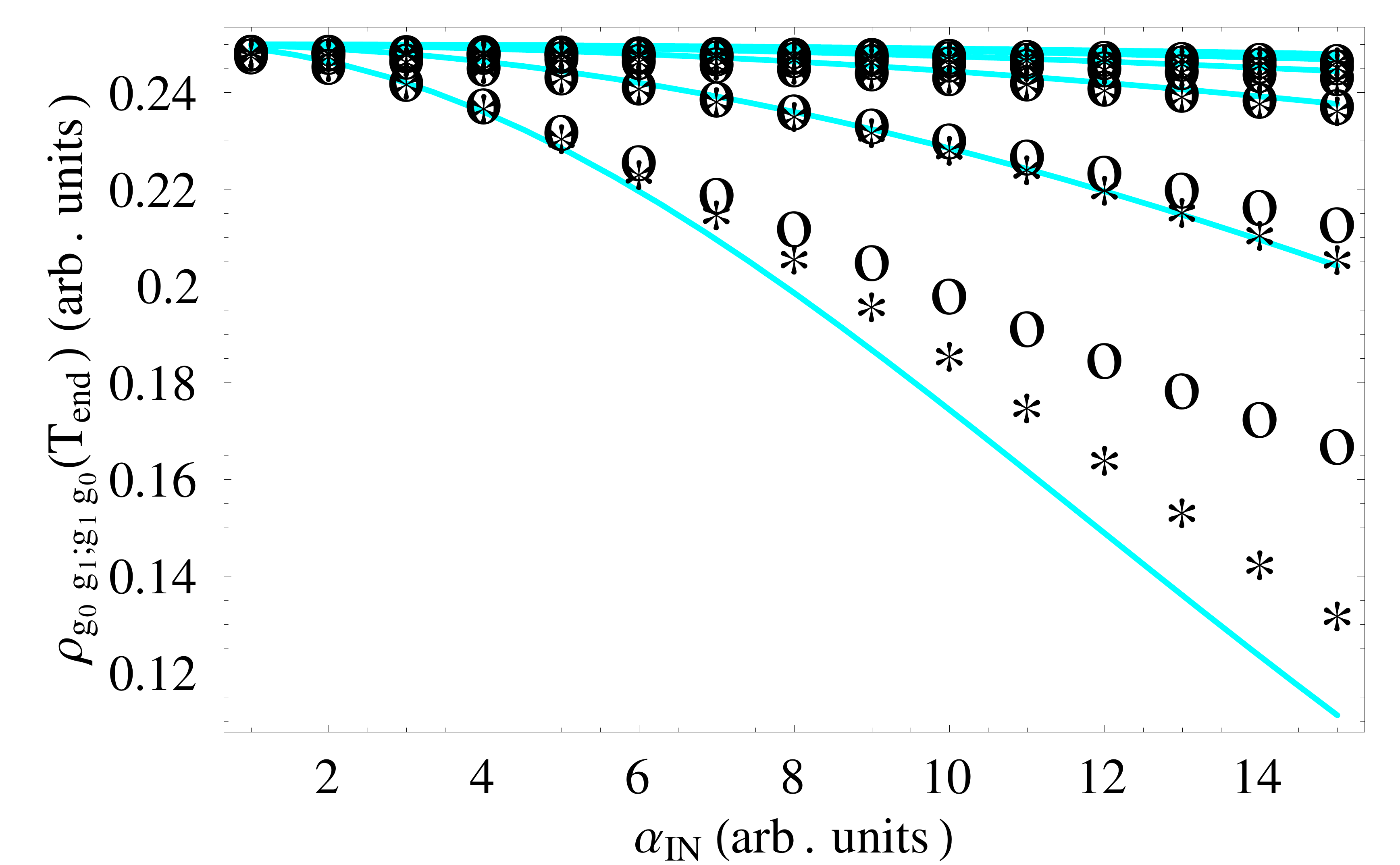}\\
(c)&(d)\\
\includegraphics[width=7cm]{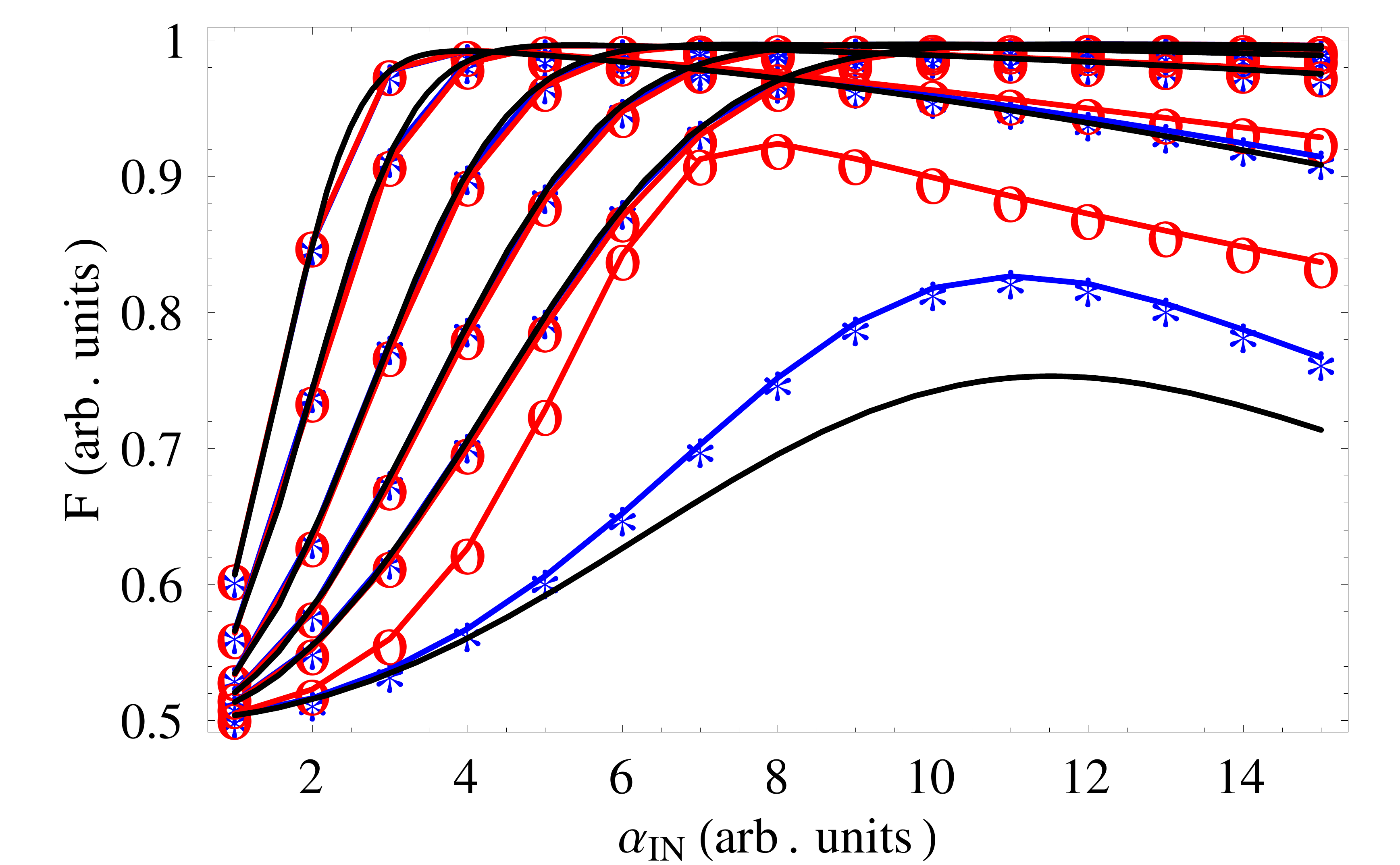}&\includegraphics[width=7cm]{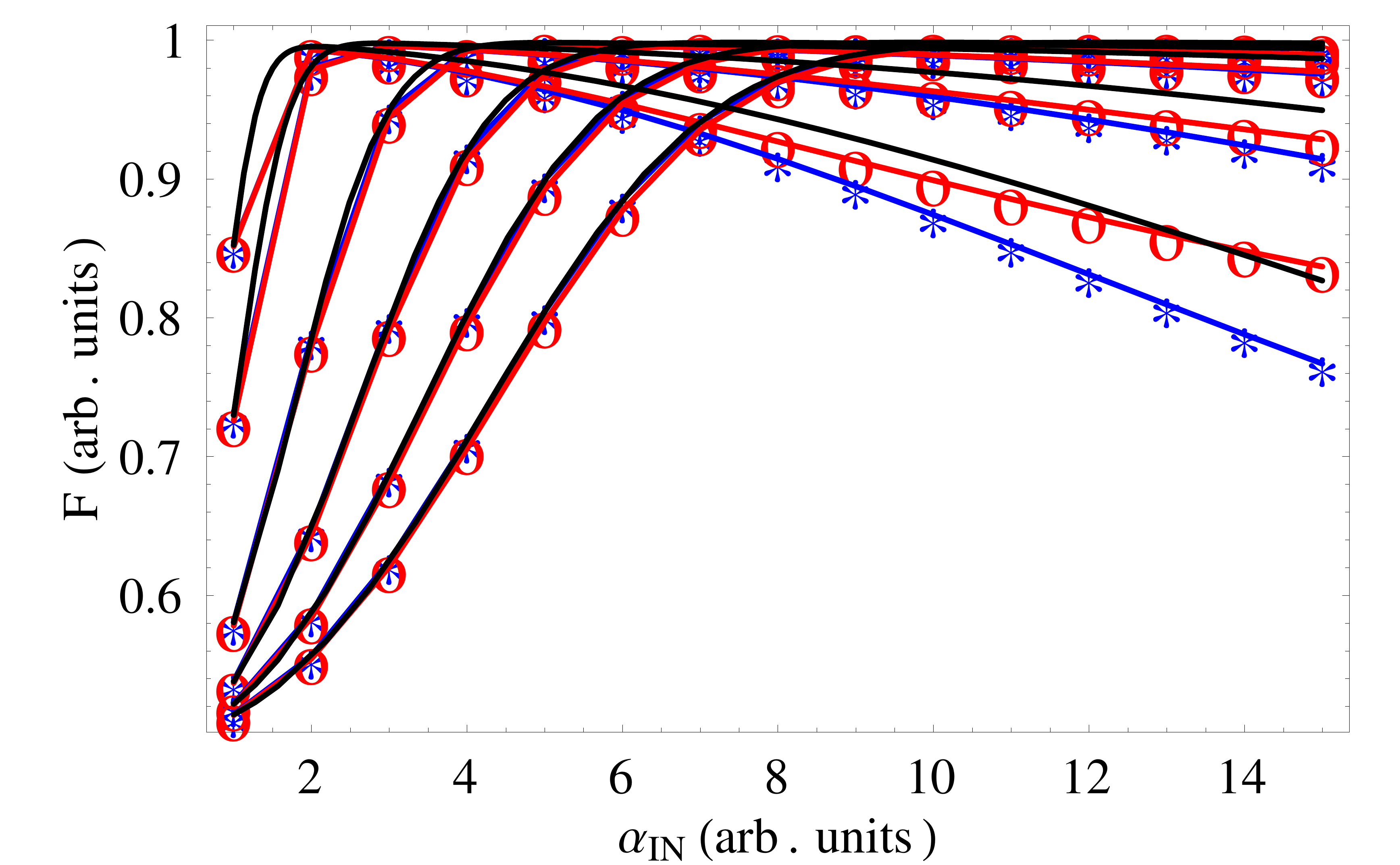}
\end{tabular}
\caption{\label{fig:semi} (Color online) Results of a semiclassical simulation.  (a) Phase shift from a simulation in terms of the approximate phase shift $arg\bigl[1/(\frac{\kappa}{2}-i\frac{g^2}{\Delta\omega_q})\bigr]$ versus $\Delta\omega$ for pulse lengths $1\;\mathrm{ns}$ (stars) and $100\;\mathrm{ps}$ (circles), and for $\alpha_{\textrm{\tiny IN}}=3$ (highest), $6,\;9,\;12$, and $15$ (lowest). For low detuning, high photon number, and short pulse length, we see a strong saturation of the phase shift. (b) Damping of coherence $\rho_{g_0 g_1;g_1 g_0}(T_{end})$ at one transition with same encoding as in (a) for $\Delta\omega=1\;\mathrm{meV}$ (lowest), $2\;\mathrm{meV},\;4\;\mathrm{meV},\;6\;\mathrm{meV},\;8\;\mathrm{meV}$, and $10\;\mathrm{meV}$ (highest). For low detuning (and thus large Stark shift) the scattering is lower compared to the approximate results from Eq. \eqref{eq:decent} (light blue line) due to reflection induced by the Stark shift. (c) Fidelity ($x_c=0.3$) for different pulse lengths (symbols) compared to the model from Sec. \ref{sec:Fid} (lines) for $\Delta\omega=1\;\mathrm{meV}$ (lowest), $2\;\mathrm{meV},\;4\;\mathrm{meV},\;6\;\mathrm{meV},\;8\;\mathrm{meV}$, and $10\;\mathrm{meV}$ (from left to right). In the high-fidelity regime, the results coincide and the large deviations for $\Delta\omega=1\;\mathrm{meV}$ are typical for very big Stark shifts $\sim\frac{\pi}{2}$ [see Eq. \eqref{eq:signal1}]. In (d), the case of tuning in between the QD resonances is displayed. From left to right the lines correspond to $|\nu_A-\nu_B|/2=1\;\mathrm{meV},\;2\;\mathrm{meV},\;4\;\mathrm{meV},\;6\;\mathrm{meV}$, and $8\;\mathrm{meV}$. The colored lines are guides to the eye.}
		
\end{figure*}

We now test the validity of the expansion in Sec. \ref{sec:Sch} by a semiclassical simulation retaining the excited states, similar to Ref. \onlinecite{Lad}. We will consider here for simplicity only the situation of identical QDs, $B_{ext}=0$, and no light holes. For the density operator of a single QD, we make a similar semiclassical ansatz as in Eq. \eqref{eq:rho}, but include also the excited states.
\small
\bea\label{eq:rhosat}
\hat{\rho}(t)&=&\sum_{q=0,1}\Bigl(\bigl(\rho_{g_q g_q}(t)|g_q\ra\la g_q|+\rho_{g_q e_q}(t)|e_q\ra\la g_q|+\rho_{e_q g_q}(t)|g_q\ra\non
&&\times\la e_q|+\rho_{e_q e_q}(t)|e_q\ra\la e_q|\bigr)\otimes|\tilde{\alpha}_q(t)\ra_q\la\tilde{\alpha}_q(t)|_q\non
&&\otimes|\alpha_{q'}(t)\ra_{q'}\la\alpha_{q'}(t)|_{q'}\Bigr)+\sum_{q=0,1}\Bigl(\bigl(\rho_{g_{q'} g_q}(t)|g_q\ra\la g_{q'}|\non
&&+\rho_{g_{q'} e_q}(t)|e_q\ra\la g_{q'}|+\rho_{e_{q'} g_q}(t)|g_q\ra\la e_{q'}|+\rho_{e_{q'} e_q}(t)|e_q\ra\non
&&\times\la e_{q'}|\bigr)\otimes|\tilde{\alpha}_q(t)\ra_q\la\alpha_q(t)|_q\otimes|\alpha_{q'}(t)\ra_{q'}\la\tilde{\alpha}_{q'}(t)|_{q'}\Bigr)\;.
\eea
\normalsize
This ansatz is based on the lowest order approximation neglecting any quantum correlations which, according to our findings in Sec. \ref{sec:Fid} and Ref. \onlinecite{Lad}, is a good approximation in the low saturation regime. First, we transform the Hamiltonian from Eq. \eqref{eq:OrigH} to an interaction picture with respect to
\be
H_0'=\sum_{q=0,1}\Bigl[\omega_0\ah_q^{\dagger}\ah_q+\frac{\nu}{2}|e_q\ra\la e_q|+(\frac{\nu}{2}-\omega_0)|g_q\ra\la g_q|\Bigr]\;,
\ee
which amounts to the replacements
\bea
&&\Sigh_q^-\rar e^{-i\omega_0 t}\Sigh_q^-\;,\quad\Sigh_q^+\rar e^{i\omega_0 t}\Sigh_q^+\;,\non
&&\ah_q\rar e^{-i\omega_0 t} \ah_q\;,
\eea
yielding for the trivial part of the Hamiltonian
\be
H_0'^{(I)}=-\sum_{q=0,1}\Delta\omega|g_q\ra\la g_q|\;.
\ee
The equations of motion that determine $\tilde{\alpha}_q(t)$, i.e., the output fields for a definite spin state $q$, are
\bea\label{eq:OBE1}
\dot{\rho}_{e_q e_q}(t)&=&i g \Bigl[\rho_{e_q g_q}(t)(\tilde{\alpha}_q(t))^*-(\rho_{e_q g_q}(t))^*\tilde{\alpha}_q(t)\Bigr]\non
&&-\Gamma\rho_{e_q e_q}(t)\;,\non
\dot{\rho}_{e_q g_q}(t)&=&i g (2\rho_{e_q e_q}(t)-\rho_{g_q g_q}(0))\\
&&-\Bigl[\frac{\Gamma}{2}+i\Delta\omega\Bigr]\rho_{e_q g_q}(t)\;,\non
\dot{\tilde{\alpha}}_q(t)&=&-\frac{\kappa}{2}\tilde{\alpha}_q(t)+\sqrt{\kappa}\frac{\alpha_{\textrm{\tiny IN}}}{\sqrt{2}}F_{\textrm{\tiny IN}}(t)-i g \rho_{e_q g_q}(t)\;,\nonu
\eea
where $\rho_{g_q g_q}(0)=1$. In Fig. \ref{fig:rho1}, we plot a typical solution to this equation for $\tau_{\textrm{\tiny P}}=100\;\mathrm{ps}$, $\alpha=4$, and $\Delta\omega=2\;\mathrm{meV}$.
The actual value for the phase shift of the light is given as an average as discussed in Sec. \ref{subsec:Cav} and the fast increase at short times displays how the steady state of the cavity is reached. The phase shifts are compared in Fig. \ref{fig:semi}(a) to the approximate result for different pulse lengths, saturating for large values of our expansion parameter. However, in the regime $\Delta\omega=2-10\;\mathrm{meV}$ the deviation is not more than $10\%$.

A test of  Eq. \eqref{eq:decent} requires, for a fixed $q$,  solving two additional equations for the simplest case of driving with circular polarized light (here, $q=0$). In this case, we have a two-level system ($|g_0\ra$, $|e_0\ra$) driven by nonresonant light. We are interested in the coherence with the other (ground state) level $|g_{1}\ra$ which does not couple for this polarization. The coherence between the two ground states, i.e., $\rho_{g_1 g_{0}}(t)$, is coupled to  $\rho_{g_1 e_{0}}(t)$ via
\bea\label{eq:OBE2}
\dot{\rho}_{g_1 g_0}(t)&=&i g \bar{\alpha}_0(t)\rho_{g_1 e_0}(t)+i\Delta\omega\rho_{g_1 g_0}(t)\;,\\
\dot{\rho}_{g_1 e_0}(t)&=&i g \bigl[\bar{\alpha}_0(t)\bigr]^*\rho_{g_1 g_0}(t)-\frac{\Gamma}{2}\rho_{g_1 e_0}(t)\;,
\eea
where $\bar{\alpha}_0(t):=\frac{1}{2}[\alpha_0(t)+\tilde{\alpha}_0(t)]$ is the solution to Eq. \eqref{eq:OBE1} for $\rho_{g_0 g_0}(0)=\rho_{g_1 g_1}(0)=\frac{1}{2}$ according to the situation of an initial equal superposition of both spin ground states. The approximate and the semiclassical results for the damping are shown in Fig. \ref{fig:semi}(b) for various detunings and pulse lengths.

For the double cavity system, the phase shifts from each cavity are added and the total decay of the entanglement coherence is determined by the contributions from each transition. Although there is, particularly for low $\Delta\omega$, an overestimation of scattering by Eq. \eqref{eq:decent}, which is due the fact that a smaller amount of light couples into the cavity due to an intensity dependent Stark shift, we encounter that for the final fidelity, the two methods are yielding practically identical  results [see Fig. \ref{fig:semi}(c)]. In Fig. \ref{fig:semi}(d), the case of tuning the laser in between the two QD resonances is shown (see also Appendix \ref{app:B} for the differences between the two cases). The saturation of the signal, lowering the fidelity, is compensated by a lower decay of the coherence. These results suggest that using the approximate model, which considerably simplifies practical calculations, is well justified.\\

Nonclassical effects have not been included so far. Going to $\mathcal{O}\bigl(\frac{g^4}{\Delta\omega^3}\bigr)$ in the expansion of Sec. \ref{subsec:Exp}, there is a term describing nonlinear dephasing (see, e.g., Ref. \onlinecite{Jos}), which is $2$ orders of magnitude smaller than the linear Stark shift. We simulated quadrature squeezing due to nonlinear dephasing and found practically no effect for parameters where the fidelity is high in Fig. \ref{fig:semi} ($\sim 0.1\%$). This corresponds to the findings in Ref. \onlinecite{Lad}. For short pulses, low detuning and high photon number squeezing may occur. In more detail, for $\Delta\omega=1\;\mathrm{meV}$ and $\alpha_{\textrm{\tiny IN}}>4$, there is squeezing on the order of a percent, whereas for higher detuning, it is negligible. Principally, squeezing would not harm entanglement creation unless one could learn about the spin state at one QD from the amount of squeezing observed. However, for our conclusions nonclassical behavior of the light is not relevant.

\section{Summary and Conclusions \label{sec:Sum}}

We have shown that entanglement with high fidelity ($>0.99$) and success probability ($>0.49$) is possible between distant quantum dot spins using cavity QED and coherent light bus modes. Nuclear spins should not matter when spin-echo techniques are used, except for the $T_2$ time they impose. A simple analytic model based on the elimination of the excited states was used, largely simplifying the determination of the optimal parameters which are necessary to achieve high fidelity. QDs are generally nonidentical in terms of transition frequency and light-matter coupling constant, but we have shown that there exist strategies which allow to largely compensate for this. Taking into account QD-specific effects, such as light-hole transitions which mainly become important if one tunes the laser in between the two QD frequencies, we demonstrate that there exist regimes where all the requirements are fulfilled.  Thus, effects from the nonidentical nature of QDs can be overcome. We mainly focused not only on one-sided cavities as they are probably most suited but also discussed two-sided cavities where long enough pulse lengths have to be used, such that the spin states are not revealed by the scattered light. We tested the simple model against a semiclassical simulation which accounts for excited state population for several pulse lengths and found it to be valid in the regime of interest. 

\begin{acknowledgements}
J.G. would like to thank D. Gerace, G. Giedke, and H. T\"ureci for helpful discussions and the Quantum Photonics group at ETH for their nice hospitality, as well as the Paul-Urban-Stipendienstiftung and the Referat f\"ur Internationale Beziehungen der Universit\"at Graz for financial support.
\end{acknowledgements}

\appendix
\begin{widetext}

\section{More general expressions for the distinguishabilities \label{app:B}}

We explicitly quote here the output states for the general case including slight detuning of the first cavity from the laser field by $\delta\omega=\omega_L-\omega_{0A}$, as used for nonidentical QDs in Sec. \ref{sec:Non}. The four different $y$-polarized output states analogous to Eq. \eqref{eq:signal1} are then $|d_{q_1 q_2}\ra=|-\frac{\alpha_{\textrm{\tiny IN}}}{2}\mathrm{Im}(\gamma^1_{q_1 q_2}-\gamma^0_{q_1q_2})\ra_y$
with the corresponding complex amplitudes $\gamma^q_{q_1 q_2}$ of the light with circular polarization $q$,
\bea
\gamma^q_{qq}&=&\Biggl(\frac{\kappa_A}{\frac{\kappa_A}{2}-i\Bigl(\delta\omega+\frac{g_A^2}{\Delta\omega_{Aq}}\Bigr)}-1\Biggr)\Biggl(\frac{\kappa_B}{\frac{\kappa_B}{2}-i\Bigl(\frac{g_B^2}{\Delta\omega_{Bq}}\Bigr)}-1\Biggr)\;,\non
\gamma^{q'}_{qq}&=&\frac{\kappa_A}{\frac{\kappa_A}{2}-i\delta\omega}-1\;,\\
\gamma^q_{qq'}&=&\frac{\kappa_A}{\frac{\kappa_A}{2}-i\Bigl(\delta\omega+\frac{g_A^2}{\Delta\omega_{Aq}}\Bigr)}-1\;,\non
\gamma^q_{q'q}&=&\Biggl(\frac{\kappa_A}{\frac{\kappa_A}{2}-i\delta\omega}-1\Biggr)\Biggl(\frac{\kappa_B}{\frac{\kappa_B}{2}-i\Bigl(\frac{g_B^2}{\Delta\omega_{Bq}}\Bigr)}-1\Biggr)\;.\nonu
\eea

\begin{figure}[t]
\includegraphics[width=7.5cm]{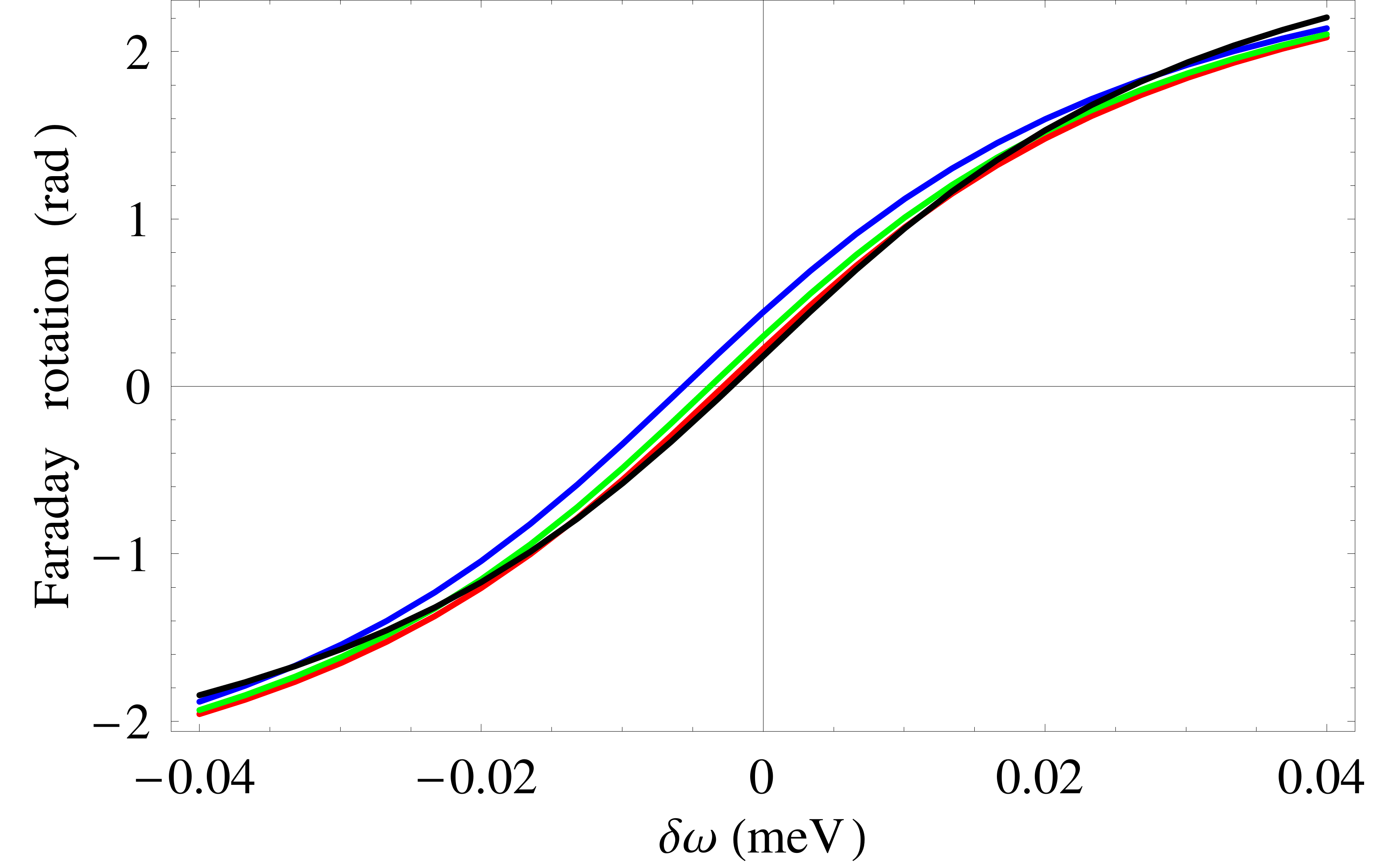}
		\caption{\label{fig:delshow} (Color online) $\mathrm{arg}(\gamma^{1}_{10})$ for $B=0$ and $\Delta\omega_{A1}=8\;\mathrm{meV}$ (red), $\Delta\omega_{A1}=6\;\mathrm{meV}$ (green), and $\Delta\omega_{A1}=4\;\mathrm{meV}$ (blue) compared to $\mathrm{arg}(\gamma^{0}_{10})$ for $\Delta\omega_{B1}=10\;\mathrm{meV}$ (black). There are two possible solutions for $\delta\omega$ where the lines intersect such that the Faraday rotation for both configurations with opposite spins is equal. }
\end{figure}


The first two expressions correspond to equal spins, while the others to opposite spins. The notion of abrupt changing cavity-laser detunings $\delta\omega$ found by numerical optimization of the fidelity (see Figs. \ref{fig:Diff1} and \ref{fig:Diff1xc15}) can be understood from the requirement that $\gamma^q_{qq'}=\gamma^{q'}_{qq'}$. As can be seen from Fig. \ref{fig:delshow}, for a fixed $|\nu_A-\nu_B|$, there are two points where this is true: one for positive $\delta\omega$ and one for negative one. The curves are clearly not symmetric in $\delta\omega$ because of the Stark shift and thus it is either more favorable to have negative or positive $\delta\omega$. As soon as the positive solution becomes more favorable there is a sudden change.  

For tuning the laser in between the two QD resonances, the amplitudes are obtained by putting a negative Stark shift at the second cavity. Thus, for the configuration with same spins, we have for identical quantum dots zero phase shifts and finite one for the other configurations.

\section{Expressions for the fidelity \label{app:A}}

The expression of the fidelity defined in Eq. \eqref{eq:fid2} gives evaluated
\bea\label{eq:fid2app}
F&=&\frac{(1/2)\int_{-x_c}^{x_c}dx\{[G_{10}(x)]^2\rho_{g_0g_0;g_1g_1}(T_{end})+[G_{01}(x)]^2\rho_{g_1g_1;g_0g_0}(T_{end})-2G_{10}(x)G_{01}(x)\mathrm{Re}[\rho_{g_0g_1;g_1g_0}(T_{end})]\}}{P_{succ}(x_c)}\non
&=&\frac{1}{8P_{succ}(x_c)}\Biggl\{\frac{1}{2} \{\mathrm{erf}[\sqrt{2}(x_c-d_{10})] + \mathrm{erf}[\sqrt{2}(x_c+d_{10})]\non
&+&\mathrm{erf}[\sqrt{2}(x_c-d_{01})] + \mathrm{erf}[\sqrt{2}(x_c+d_{01})]\}\non
&+&e^{-(1/2)(d_{10}-d_{01})^2}\Biggl[\mathrm{erf}\Biggl(\frac{d_{10}+d_{01}+2x_c}{\sqrt{2}}\Biggr)-\mathrm{erf}\Biggl(\frac{d_{10}+d_{01}-2x_c}{\sqrt{2}}\Biggr)\Biggr]\;e^{-(\alpha_{\textrm{\tiny IN}}^2/2)}\sum_x[(\Gamma_{x0}^R+\Gamma_{x1}^R)/2]\Phi_x\Biggr\}\;,
\eea
with the success probability
\bea
P_{succ}(x_c)&=&\int_{-x_c}^{x_c}dx \{[G_{11}(x)]^2\rho_{g_1g_1;g_1g_1}(T_{end})+[G_{00}(x)]^2\rho_{g_0g_0;g_0g_0}(T_{end})\non
&&+[G_{10}(x)]^2\rho_{g_0g_0;g_1g_1}(T_{end})+[G_{01}(x)]^2\rho_{g_1g_1;g_0g_0}(T_{end})\}\non
&=&\frac{1}{8}\{\mathrm{erf}[\sqrt{2}(x_c-d_{11})] + \mathrm{erf}[\sqrt{2}(x_c+d_{11})] + \mathrm{erf}[\sqrt{2}(x_c-d_{00})]\non 
&&+ \mathrm{erf}[\sqrt{2}(x_c+d_{00})] + \mathrm{erf}[\sqrt{2}(x_c-d_{10})] + \mathrm{erf}[\sqrt{2}(x_c+d_{10})]\non 
&&+ \mathrm{erf}[\sqrt{2}(x_c-d_{01})] + \mathrm{erf}[\sqrt{2}(x_c+d_{01})]\}\;.
\eea

For the two-sided cavity scenario, the ansatz for the density operator [cf. Eq. \eqref{eq:rho} for the one-sided case] contains the photons transmitted and reflected from the double-cavity system. Using Eq. \eqref{eq:TS} for the reflected light when the spins are opposite, we obtain after tracing out the reflected light the fidelity by replacing in Eq. \eqref{eq:fid2app}
\be
\mathrm{Re}[\rho_{g_0g_1;g_1g_0}(T_{end})]\rightarrow \mathrm{Re}[\rho_{g_0g_1;g_1g_0}(T_{end})\la \beta^{g_0 g_1}|_{yL} |\beta^{g_1 g_0}\ra_{yL}]\;,
\ee
with
\be
\la \beta^{g_0 g_1}|_{yL} |\beta^{g_1 g_0}\ra_{yL}=e^{-(1/2)[\sum_{xq} \tilde{d}_{xq}^2-2 I_{ol}(\tau_{\textrm{\tiny P}})(\tilde{d}_{A1}\tilde{d}_{B0}+\tilde{d}_{A0}\tilde{d}_{B1})]-\tilde{d}_{A0} \tilde{d}_{A1}-\tilde{d}_{B0} \tilde{d}_{B1}+I_{ol}(\tau_{\textrm{\tiny P}})(\tilde{d}_{A0}\tilde{d}_{B0}+\tilde{d}_{B1}\tilde{d}_{A1})}\;,
\ee
where $\tilde{d}_{xq}$ and $I_{ol}(\tau_{\textrm{\tiny P}})$ are defined as in Sec. \ref{subsec:Dou}.
\end{widetext}

\end{document}